\documentclass[useAMS,usenatbib,usegraphicx]{mn2e}
\usepackage{amssymb}
\usepackage{amsmath}
\usepackage{color} 
\newcommand{\eqd}{\,\, .}
\newcommand{\eqc}{\,\, ,}
\newcommand{\pd}[1]{\, \partial #1 \,}
\newcommand{\td}[1]{\, \mathrm{d} #1 \,}
\newcommand{\intl}{\int\limits}

\newcommand{\HF}[1]{\; \mathrm{H}\left[ #1 \right]}  
\newcommand{\DF}[1]{\; \delta\left( #1 \right)}  
\newcommand{\lec}{(1+l_{ec})}
\newcommand{\Gamec}{\Gamma_{ec}}
\newcommand{\eps}{\epsilon}
\newcommand{\epss}{\epsilon_1}
\newcommand{\epsn}{\epsilon_0}
\newcommand{\epst}{\epsilon_{t}}
\newcommand{\epec}{\epsilon_{ec}}
\newcommand{\epsmax}{\epsilon_{1,max}}
\newcommand{\epk}[1]{\epsilon_{pk, #1 }}
\newcommand{\epstra}[1]{\epsilon_{tra, #1}}
\newcommand{\gamf}{\gamma_{0,4}}
\newcommand{\taug}{\tau_{\gamma\gamma}}
\newcommand{\taum}[1]{\tau_{max, #1 }}
\newcommand{\taus}{\tau_{ssa}}
\title[Optical Depth in blazars in a non-linear scenario]{Blazars and Optical Depth in a non-linear, time-dependent injection and cooling scenario}
\author[M. Zacharias]{M. Zacharias \\
Landessternwarte, Universit\"at Heidelberg, K\"onigstuhl, D-69117 Heidelberg, Germany \\
m.zacharias@lsw.uni-heidelberg.de}
\date{Received ?; accepted ? }
\begin{document}
\maketitle
\begin{abstract}
In this paper the optical depths in blazars due to photo-pair production is calculated for a time-dependent, non-linear injection model. Several target photon fields are taken into account, namely the internal synchrotron, synchrotron-self Compton and external Compton radiation, as well as a constant external soft photon field. By applying the optical depths to theoretical blazar spectra only the constant external photon field turns out to significantly influence the radiation at high energies. The impact of the internal time-dependent radiation fields is either minor or requires extreme parameter settings. Additionally, the synchrotron-self absorption turn-over energy for low synchrotron energies is calculated, which is inherently time-dependent. It would be challenging to use it to constrain free parameters, since precise knowledge of the observation time relative to the injection time is needed. In conclusion, optical depth does not significantly influence the non-linear, time-dependent injection and cooling model. 
\end{abstract}
\begin{keywords}
radiation mechanisms: non-thermal -- BL Lacertae objects: general -- galaxies: active -- relativistic processes
\end{keywords}
%
%
\section{Introduction} \label{sec:intro}
In view of the unified model of active galactic nuclei \citep{up95} blazars are active galaxies, where the angle between the line of sight and the jet is very small and the emission of the jet is strongly Doppler boosted outshining in some cases the host galaxy. Blazars are variable on all time scales from years down to just a few minutes. Thus, they are extremely useful to analyse the properties and physics of relativistic jets in very different emission states. 

The spectral energy distribution (SED) of blazars is characterized by two broad components. The low energetic one peaks in the infrared to X-ray part of the spectrum, and is attributed to synchrotron emission of highly relativistic electrons. The high energetic component peaks in the MeV to TeV energy regime, and its origin is a matter of debate. In leptonic models \citep{b07} the high energetic emission is attributed to inverse Compton radiation by the same electron population scattering the ambient photon fields, namely the self-made synchrotron emission (synchrotron-self Compton, SSC, \citet{jos74}) or ambient external photon fields (external Compton, EC, e.g. \citet{ds93,sbr94,bea00}). In hadronic models the high energetic component originates from protons, either by direct proton synchrotron emission, or via by-products of photo-meson production \citep{m93,bea13,cea14}. This paper deals with the leptonic scenario, where protons, if available, only serve as a cold background.

The standard one-zone model for blazar emission attributes the detected radiation to a homogeneous, spherical volume located somewhere in the jet. The particles are continuously injected into the blob resulting in an equilibrium situation, which has the advantage of an easy mathematical solution for the differential equation describing the electron distribution function. This model has been successfully used in many blazars to describe both quiescence and flaring modes.

In recent publications a small change of the one-zone model has been discussed, which has profound implications for the resulting spectra and light curves. Namely, the continuous injection has been replaced by a time-dependent injection.

In such a scenario the electrons cannot reach equilibrium, as long as no continuous re-acceleration is provided. Since the electron cooling due to SSC emission depends on the self-produced synchrotron photon energy density, which in turn depends on the electron distribution function itself, the SSC electron cooling becomes non-linear and time-dependent. Obviously, this has strong consequences. As was shown analytically by \cite{s09} and \cite{zs10} using an instantaneous injection of the radiating particles at time $t=0$, the non-linear cooling significantly reduces the cooling time scales compared to the standard linear (and continuous) model. 

Additionally, in a non-equilibrium model the cooling behaviour can change with severe impacts on the emerging radiation. On the one hand, cooling due to synchrotron and external Compton emission is linear, at least in a simple model as utilised in this work (but see for example \citet{sl07}). On the other hand, the efficiency of the SSC cooling term decreases with respect to time, since the energy density in the synchrotron photons decreases due to the energy losses of the electrons. Hence, after some time the linear cooling terms become stronger than the SSC cooling term, and the cooling behaviour changes from non-linear to linear. Using the instructive example of a monochromatic particle injection with initial electron Lorentz factor $\gamma=\gamma_0$, the changing cooling effect has been explored first by \cite{sbm10}. Further implications have been presented by \cite{zs12a,zs12b,zs13,z14}. In summary, each component of the emerging SED exhibits an additional break, which is solely due to the changing cooling behaviour without the need for complicated electron distributions. Furthermore, the emerging lightcurves show strong differences compared to a purely linear model, like a reduction of the variability time scale at high energies, while the variability at low energies follows the conventional time scales.

In this work the important aspect of photo-pair production in the time-dependent injection scenario is discussed. In photo-pair production a high energetic photon interacts with a low energetic photon producing an electron-positron pair. If the energy sum of both photons exceeds at least twice the electron rest energy an electron-positron pair is created, while the photons are destroyed. Hence, the source becomes optically thick and the detected flux can be significantly reduced. Here, the implications on the optical depth by the time-dependent injection model are presented and compared to the standard one-zone model. 

In section \ref{sec:gamgamabs} the photo-pair production and the necessary integral formula is described. The following sections introduce first the soft photon fields, namely the internal time-dependent synchrotron, SSC and EC radiation, followed by the calculation of the respective optical depths. In section \ref{sec:ext} the optical depth due to the external soft photon field is calculated. The influence of all these photon fields on the SED is discussed in section \ref{sec:sed} by applying the absorption to model SEDs. The results are summarized in section \ref{sec:discon}.

Synchrotron-self absorption is an important process for low energetic synchrotron photons. The calculations have been performed by \cite[ch. 3]{z13phd} and are briefly described in appendix \ref{app:ssa} for the sake of completeness. The description concentrates on the synchrotron-self absorption transition energy, which can potentially be used to constrain the free parameters. Details with respect to the emerging synchrotron SED can be found in \cite{z13phd}.
%
%
\section{Photo-pair production} \label{sec:gamgamabs}
High energy photons can be absorbed in photo-pair production where the high energy photon couples with a low energetic target photon producing an electron-positron pair:
\begin{align*}
\gamma + \gamma \rightarrow e^{-} + e^{+} \eqd
\end{align*}
The optical depth is calculated along the path of the $\gamma$-ray from its origin to the observer through a target photon field. If the target photon field is isotropically distributed in a sphere around the $\gamma$-ray radiation zone the path length equals the radius $R$ of the target photon sphere. Then, the optical depth is derived by 
\begin{align}
\taug(\epss,t) = R \intl_0^{\infty} n_{ph}(\eps,t) \sigma_{\gamma\gamma}(\eps,\epss) \td{\eps} \label{eq:taugggen1} \eqc
\end{align}
with the target photon number density $n_{ph}(\eps,t)$, the pair production cross section $\sigma_{\gamma\gamma}(\eps,\epss)$, the normalized target photon energy $\eps$, and the normalized energy of the high energetic photons $\epss$. The energies are normalized to the electron rest energy as $\eps = E_{ph}/mc^2$. For internal target photon fields the path length $R$ equals the radius of the emission blob $R_b$, while for external soft target photons the path length is denoted as $R_{ex}$ being typically larger than $R_b$. The time variable is marked as $t$.

Since the intensity of an optically thick source is corrected by the factor $e^{-\taug}$, a rule of thumb says that a source is opaque if $\taug(\epss,t)>1$. This rule is utilzed below to derive the energy regime for which the source is optically thick.

The pair production cross section is given by, e.g., \cite{aea08} as
\begin{align}
\sigma_{\gamma\gamma}(s) =& \frac{3\sigma_T}{2s^2} \left[\left( s+\frac{1}{2}\ln{s} - \frac{1}{6} + \frac{1}{2s} \right) \ln{\left( \sqrt{s}+\sqrt{s-1} \right)} \right. \nonumber \\
& - \left. \left( s+\frac{4}{9}-\frac{1}{9s} \right) \sqrt{1-\frac{1}{s}} \right] \label{eq:sigmagggen1} \eqc
\end{align}
with $s=\eps\epss$, and the Thomson cross section $\sigma_T = 6.65\cdot 10^{-25}$ cm$^2$.

This form of the cross section is not useful for analytical calculations. A simpler form of the cross section is used below, which takes into account that the maximum of the cross section is actually very close to the minimal energy requirement $\eps = 2/\epss$ (in order to have sufficient energy to produce the electron-positron pair at rest). Hence, a delta-function approximation is employed, which is according to \citet[ch. 10.2]{dm09}
\begin{align}
\sigma_{\gamma\gamma}(\eps,\epss) = \frac{\sigma_T}{3}\eps \DF{\eps-\frac{2}{\epss}} \label{eq:sigmagggen2} \eqd
\end{align}

The target photon number density $n_{ph}(\eps,t)$ can be related to the target photon intensities $I(\eps,t)$ as follows:
\begin{align}
n_{ph}(\eps,t) = \frac{I(\eps,t)}{c\eps} \label{eq:photdens1} \eqc
\end{align}
with the speed of light $c = 3\cdot 10^{10}$ cm/s.

Inserting equations (\ref{eq:sigmagggen2}) and (\ref{eq:photdens1}) in equation (\ref{eq:taugggen1}) yields
\begin{align}
\taug(\epss,t) = \frac{R\sigma_T}{3c} I\left( \eps=\frac{2}{\epss} , t \right) \label{eq:tauggint1} \eqc
\end{align}
where the path length $R$ needs to be specified according to the cases considered.

Below, the optical depth is calculated for several target photon fields. First, synchrotron and SSC target photons are utilised (section \ref{sec:synssc}), followed by the EC photons (section \ref{sec:ec}). The respective intensities necessary for equation (\ref{eq:photdens1}) are presented in section \ref{sec:intdis}, and briefly derived in appendix \ref{app:int}. For these three cases $R=R_b$. 

As a fourth example the external photons themselves are used, although the target photon number density needs to be derived in a slightly different way, which is described in section \ref{sec:ext}. In this case the path length becomes $R=R_{ex}$.
%
%
\section{Internal intensity distributions} \label{sec:intdis}
The derivation of the internal synchrotron, SSC and EC intensities in the outlined scenario (i.e., instantaneous injection of monochromatic electrons: $Q(\gamma,t)\propto \DF{\gamma-\gamma_0}\DF{t}$) has been performed by \cite{sbm10,zs12a,zs12b}. A short summary of the calculations and the definitions and values of the parameters is given in appendix \ref{app:int}. In this section, only the important results for the following calculations are quoted.

The solutions to the kinetic equation (see equation (\ref{eq:kineq1})) are dependent on two parameters, namely the external Compton parameter $l_{ec}$ and the injection parameter $\alpha$. They are discussed in some detail, since they govern the cooling behaviour of the source. 

The external Compton parameter $l_{ec}$ is defined as the ratio of the linear cooling terms:
\begin{align}
l_{ec} &= \frac{|\dot{\gamma}_{ec}|}{|\dot{\gamma}_{syn}|} \nonumber \\
&= \frac{4\Gamma_b^2}{3} \frac{u_{ec}^{\prime}}{u_B} \label{eq:deflec} \eqd
\end{align}
Here, $u_B=B^2/8\pi$ is the magnetic energy density of a tangled magnetic field with strength $B=b$Gauss,\footnote{Note the result calculated by \cite{b14} that for polarized emission, i.e. in an ordered magnetic field, the photo-pair production is less important than in a tangled magnetic field.} and $u_{ec}^{\prime}$ the energy density in external photons in the galactic frame. The prefactor gives the angle-averaged Lorentz transformation from the galactic to the comoving frame for an isotropic photon field traversing the emission blob. The blob moves with Lorentz factor $\Gamma_b$ \citep{ds93}. 

The injection parameter $\alpha$ is defined as the square root of the ratio of the non-linear to the linear cooling terms at time of injection:
\begin{align}
\alpha^2 &= \frac{|\dot{\gamma}_{ssc}(t=0)|}{|\dot{\gamma}_{syn}|+|\dot{\gamma}_{ec}|} \nonumber \\
&= \frac{A_0Q_0\gamma_0^2}{D_0\lec} \label{eq:defalpha} \eqd
\end{align}
The parameters are $D_0 = 1.256\cdot 10^{-9}b^2$ s$^{-1}$, and $A_0 = 1.15\cdot 10^{-18} R_{15}b^2$ cm$^{3}$s$^{-1}$. The initial Lorentz factor of the electrons is $\gamma_0$. The source radius of $R_b=10^{15}R_{15}$ cm is scaled according to the rule $K = 10^x K_x$ in cgs-units. The electron density is given by $Q_0$.

For $\alpha<1$ the cooling is dominated by the linear cooling terms implying linear cooling for all times. If $\alpha>1$, the non-linear cooling term initially dominates for times $x<x_c$, where the convenient normalized time $x=D_0\lec\gamma_0 t$ is introduced. For later times $x>x_c$ the linear cooling terms control the cooling behaviour. The normalized cross-over time is defined as 
\begin{align}
x_c=\frac{\alpha^3-1}{3\alpha^2} \label{eq:defxc}
\end{align} 
depending solely on the value of the injection parameter. The change in the cooling behaviour is the important consequence of the model having strong consequences for the resulting SEDs and lightcurves, as is outlined in the introduction. 

Equation (\ref{eq:defalpha}) implies that a higher density $Q_0$ of particles increases the probability of non-linear cooling, while a strong external source ($l_{ec}\gg 1$) decreases this probability. A large value of the injection parameter could be realised by a small emission region, since $\alpha\propto R_b^{-1}$ (see also equation (19) of \cite{sbm10}).

Both equations (\ref{eq:deflec}) and (\ref{eq:defalpha}) are strictly valid only in the Thomson regime of inverse Compton scattering, since the cooling terms as given in appendix \ref{app:int} neglect the influence of the Klein-Nishina effect at high electron energies. The inclusion of Klein-Nishina effects gives interesting results, such as a hardening of the emitted synchrotron spectrum at high energies or the suppression of inverse Compton radiation at high $\gamma$-ray energies \citep{bea97,da02,mea05}. However, the inclusion of the Klein-Nishina cooling term in the non-linear model would complicate the calculations by a lot, which is the reason why most of the studies are done numerically even for equilibrium models. For simplicity and ease of the analytical work, the Klein-Nishina cooling is neglected in the cooling terms. It is, however, included in the calculation of the SSC and external Compton intensity in appendix \ref{app:intssc} and \ref{app:intec}, respectively.

The soft external photon field, which is necessary to calculate the intensity of external Compton photons, is modelled as a line-like $\delta$-function. Hence, the differential external photon density transformed to the blob frame is given by
\begin{align}
u_{ec}(\eps) = \frac{4\Gamma_b^2}{3} u_{ec}^{\prime} \DF{\eps-\epec} \label{eq:defuecline} \eqc
\end{align} 
with the normalized line energy $\epec$ \citep{zs12b}.

Performing the steps outlined in appendix \ref{app:int}, the synchrotron, SSC, and EC intensities become for $\alpha<1$, respectively:
\begin{align}
I_{syn}(\eps,x) = & I_{syn,0} \left( \frac{\eps}{\epsn} \right)^{1/3} \left( 1+x \right)^{2/3} e^{-\frac{\eps}{\epsn}\left( 1+x \right)^2} \label{eq:intsyn0} \eqc \\
I_{ssc}(\eps,x) = & I_{ssc,0} \left( \frac{\eps}{\epst} \right)^{1/3} \left( 1+x \right)^{4/3} e^{-\frac{\eps}{\epst}\left( 1+x \right)^4} \nonumber \\
&\times \HF{\frac{\gamma_0}{1+x}-\eps} \label{eq:intssc0} \eqc \\
I_{ec}(\eps,x)  = & I_{ec,0} \eps \left( 1+x \right)^2 G(q_0(\eps,\epec,x)) \nonumber \\
&\times \HF{\frac{\gamma_0}{1+x}-\gamma_{min}(\epec)} \label{eq:intec0} \eqd
\end{align}
Here, $\HF{w}$ denotes Heaviside's step function with $\HF{w\geq 0}=1$, and $\HF{w<0}=0$.

For $\alpha>1$ the intensities are split at the normalized time $x=x_c$, yielding for smaller times
\begin{align}
I_{syn}(\eps,x<x_c) = & I_{syn,0} \left( \frac{\eps}{\epsn} \right)^{1/3} \left( 1+3\alpha^2 x \right)^{2/9} \nonumber \\
&\times e^{-\frac{\eps}{\epsn}\left( 1+3\alpha^2 x \right)^{2/3}} \label{eq:intsyn1} \eqc \\
I_{ssc}(\eps,x<x_c) = & I_{ssc,0} \left( \frac{\eps}{\epst} \right)^{1/3} \left( 1+3\alpha^2 x \right)^{4/9} \nonumber \\
&\times e^{-\frac{\eps}{\epst}\left( 1+3\alpha^2 x \right)^{4/3}} \HF{\frac{\gamma_0}{\left( 1+3\alpha^2 x \right)^{1/3}}-\eps} \label{eq:intssc1} \eqc \\
I_{ec}(\eps,x<x_c)  = & I_{ec,0} \eps \left( 1+3\alpha^2 x \right)^{2/3} G(q_1(\eps,\epec,x)) \nonumber \\
&\times \HF{\frac{\gamma_0}{\left( 1+3\alpha^2 x \right)^{1/3}}-\gamma_{min}(\epec)} \label{eq:intec1} \eqc
\end{align}
respectively. For larger times the intensities become
\begin{align}
I_{syn}(\eps,x>x_c) = & I_{syn,0} \left( \frac{\eps}{\epsn} \right)^{1/3} \left( \alpha_g+x \right)^{2/3} \nonumber \\
&\times e^{-\frac{\eps}{\epsn}\left( \alpha_g+x \right)^2} \label{eq:intsyn2} \eqc \\
I_{ssc}(\eps,x>x_c) = & I_{ssc,0} \left( \frac{\eps}{\epst} \right)^{1/3} \left( \alpha_g+x \right)^{4/3} \nonumber \\
&\times e^{-\frac{\eps}{\epst}\left( \alpha_g+x \right)^4} \HF{\frac{\gamma_0}{\alpha_g+x}-\eps} \label{eq:intssc2} \eqc \\
I_{ec}(\eps,x>x_c)  = & I_{ec,0} \eps \left( \alpha_g+x \right)^2 G(q_2(\eps,\epec,x)) \nonumber \\
&\times \HF{\frac{\gamma_0}{\alpha_g+x}-\gamma_{min}(\epec)} \label{eq:intec2} \eqc
\end{align}
respectively.

The constants are 
\begin{align}
I_{syn,0} &= 5.57\cdot 10^{19} \frac{\alpha^2\lec b}{\gamf^2} \; \frac{\mathrm{erg}}{\mathrm{cm}^2\mathrm{s}\;\mathrm{erg}} \label{eq:defIsyn0} \eqc \\
I_{ssc,0} &= 3.67\cdot 10^{12} \frac{\alpha^4\lec^2b}{\gamf^4}  \; \frac{\mathrm{erg}}{\mathrm{cm}^2\mathrm{s}\;\mathrm{erg}} \label{eq:defIssc0} \eqc \\
I_{ec,0} &= 6.3\cdot 10^{7} \frac{\alpha^2\lec l_{ec} b^2}{\eps_{ec,-5}^2 \gamf^4} \; \frac{\mathrm{erg}}{\mathrm{cm}^2\mathrm{s}\;\mathrm{erg}} \label{eq:defIec0} \eqc
\end{align}
and $\alpha_g=(1+2\alpha^3)/3\alpha^2$. The characteristic normalized synchrotron photon energy is given by $\epsn = 3.4\cdot 10^{-6}b\gamf^2$, while the normalized SSC-Thomson energy equals $\epst = 1.4\cdot 10^3 b\gamf^4$, and $\gamf = \gamma_0 / 10^{4}$.

The definition of the function $G(q_i(\eps,\epec,x))$ is deferred to section \ref{sec:ec}, equations (\ref{eq:Gqi1}) to (\ref{eq:gamec1}). The remaining definitions and parameter values can be found in appendix \ref{app:int}.

%
%
\section{Photo-pair production for synchrotron and synchrotron-self Compton photon targets} \label{sec:synssc}
As can be seen from equations (\ref{eq:intsyn0}) to (\ref{eq:intec2}) in section \ref{sec:intdis}, the synchrotron and the SSC intensities exhibit a similar structure, which can be summarized as follows:
\begin{align}
I_i(\eps,x) = I_{i,0} \left( \frac{\eps}{\eps_i} \right)^{1/3} \left( v+px \right)^{n/3} e^{-\frac{\eps}{\eps_i}\left( v+px \right)^n} \label{eq:Issgen} \eqd
\end{align}
The values for $I_{i,0}$, $\eps_i$, $v$, $p$, and $n$ for the respective cases can be deduced from section \ref{sec:intdis} and appendix \ref{app:int}.

With the general equation (\ref{eq:Issgen}) the optical depth due to synchrotron and SSC radiation can be calculated in one step. From equation (\ref{eq:tauggint1}) one obtains
\begin{align}
\taug(\epss,x) = \tau_i \left( \frac{\epss}{\epk{i}} \right)^{-1/3} \left( v+px \right)^{n/3} e^{-\left( \frac{\epss}{\epk{i}} \right)^{-1}\left( v+px \right)^n} \label{eq:tausynssc1} \eqc
\end{align} 
where $\tau_i = I_{i,0}R_b\sigma_T / 3c$, and $\epk{i} = 2/\eps_i$. The energy of maximum optical depth becomes
\begin{align}
\epsmax(x) = 3\epk{i} \left( v+px \right)^n \eqc \label{eq:emaxsynssc1}
\end{align}
implying an increase of the maximum energy with respect to time. This is a first indication that the optical depth of the source is not constant for these two processes. Inserting equation (\ref{eq:emaxsynssc1}) into equation (\ref{eq:tausynssc1}) yields the maximum value of the optical depth:
\begin{align}
\taum{i} = \tau_i \left( \frac{1}{3} \right)^{1/3} e^{-1/3} \eqd \label{eq:taumaxsynssc1}
\end{align}
Using the values for the synchrotron and SSC intensity the maximum values become, respectively:
\begin{align}
\taum{syn} =& 0.2 \frac{\alpha^2 b R_{15} \lec}{\gamf^2} \label{eq:taumaxsyn} \eqc \\
\taum{ssc} =& 1.4\cdot 10^{-8} \frac{\alpha^4 b R_{15} \lec^2}{\gamf^4} \label{eq:taumaxssc} \eqd
\end{align}

Having found a specific maximum value for the optical depth implies that there should be two solutions for the equation $\taug(\epstra{i},x)=1$, where $\epstra{i}$ is the transition energy from the optically thin regime to the optically thick regime. Since equation (\ref{eq:tausynssc1}) can be divided into two parts (depending on the value of the exponential function), the transition energies for both parts can be found as follows. Assuming at first
\begin{align}
\left( \frac{\epss}{\epk{i}} \right)^{-1}\left( v+px \right)^n \ll 1
\end{align}
the exponential can be neglected, giving
\begin{align}
\taug(\epstra,x) \approx& \tau_i \left( \frac{\epstra{i}}{\epk{i}} \right)^{-1/3} \left( v+px \right)^{n/3} \nonumber \\ 
\stackrel{!}{=}& 1 \nonumber \eqd
\end{align}
Hence,
\begin{align}
\epstra{i1}(x) = \tau_i^3 \epk{i} \left( v+px \right)^n \label{eq:epstrasynssc1} \eqd
\end{align}
The other part is defined by
\begin{align}
\left( \frac{\epss}{\epk{i}} \right)^{-1}\left( v+px \right)^n \gg 1 \eqd
\end{align}
Slightly rewriting equation (\ref{eq:tausynssc1}), it can be approximated as
\begin{align}
\taug(\epss,x) =& \tau_i e^{-\left( \frac{\epss}{\epk{i}} \right)^{-1}\left( v+px \right)^n + \frac{1}{3}\ln{[\left( \frac{\epss}{\epk{i}} \right)^{-1} \left( v+px \right)^{n}]}} \nonumber \\
\approx& \tau_i e^{-\left( \frac{\epss}{\epk{i}} \right)^{-1}\left( v+px \right)^n} \eqd
\end{align}
Equalling this to unity, the transition energy becomes
\begin{align}
\epstra{i2}(x) = \frac{\epk{i}}{\ln{\tau_i}}\left( v+px \right)^n \label{eq:epstrasynssc2} \eqd
\end{align}

Since the transition energies imply 
\begin{align}
\epstra{i2}(x)<\epss<\epstra{i1}(x) \label{eq:transrangesynssc1}
\end{align}
for the optically thick regime, the relation $\tau_i^3 \ln{\tau_i} > 1$ must hold. Otherwise, there is no absorption. This is fulfilled for values $\tau_i\gtrsim 1.4$, which can be used to constrain $\alpha$ in the synchrotron case
\begin{align}
\alpha > 1.8 \frac{\gamf}{b^{1/2} R_{15}^{1/2} \lec^{1/2}} \label{eq:alphasyn1} \eqc
\end{align}
and in the SSC case
\begin{align}
\alpha > 84.8 \frac{\gamf}{b^{1/4} R_{15}^{1/4} \lec^{1/2}} \label{eq:alphassc1} \eqd
\end{align}
The optical depths due to synchrotron and SSC target photon fields cannot be neglected in non-linearly cooling sources with $\alpha\gg 1$. For linearly cooling sources it depends strongly on the source parameters if photo-pair production by a synchrotron target must be taken into account.

The interesting consequence of the limited range of energies, which is affected by the optical depth according to equation (\ref{eq:transrangesynssc1}), is that not all photons above a certain threshold are actually captured. High energy photons might still escape, while lower energy photons cannot. Over time both limits increase. Interestingly, since the intensities drop significantly faster for high energies than for low energies, no more high energy photons might be produced when the source becomes optically thick for them. The lower energy photons are in the optical thick regime at first and only at later times are able to escape. This is a completely different behaviour compared to cases, where the optical depth does not change over time (e.g., for a steady external photon source as in section \ref{sec:ext}), and a direct consequence of the time-dependent injection. 

\begin{figure}
	\centering
		\includegraphics[width=0.48\textwidth]{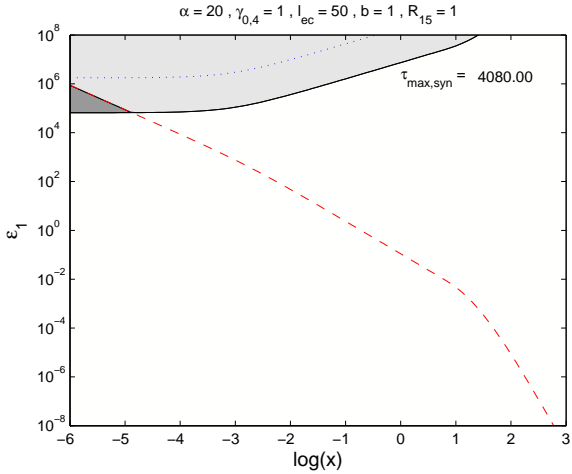}
	\caption{The plot displays the photon energy $\epss$ versus time. The shaded area shows the optically thick energy range due to interactions with a synchrotron target field. The darker area shows the range where the optical depth actually applies, since the red dashed curve shows the cut-off energy of the SSC intensity at the respective time. The blue dotted line marks $\epsmax(x)$, with its optical depth value $\taum{syn}$ given in the top right corner. Parameters are given at the top.}
	\label{fig:syn}
\end{figure}
\begin{figure}
	\centering
		\includegraphics[width=0.48\textwidth]{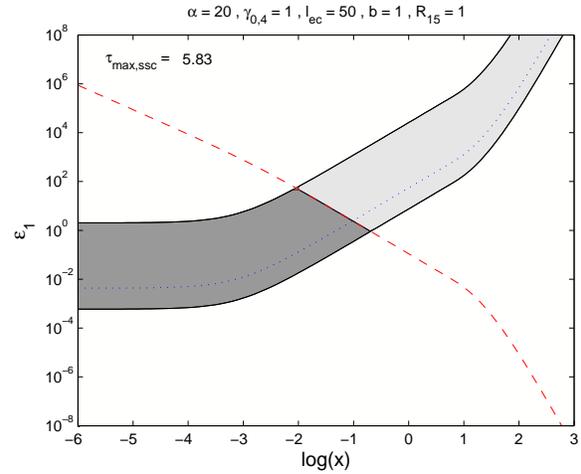}
	\caption{The plot displays the photon energy $\epss$ versus time. The shaded area shows the optically thick energy range due to interactions with an SSC target field. The darker area shows the range where the optical depth actually applies, since the red dashed curve shows the cut-off energy of the SSC intensity at the respective time. The blue dotted line marks $\epsmax(x)$, with its optical depth value $\taum{ssc}$ given in the top left corner. Parameters are given at the top.}
	\label{fig:ssc}
\end{figure}

Figures \ref{fig:syn} and \ref{fig:ssc} show the described behaviour of the optical depth for both synchrotron and SSC photons as targets, respectively. The figures are an energy ($\epss$) versus time ($x$) plot, where the energy range, which is subject to absorption by the target photons, is grey shaded. The red dashed curve shows the cut-off energy of the SSC intensity, which serves as an example for the time-dependent behaviour of the high energy photons (see appendix \ref{app:ssccoe}). SSC photons of a given energy are emitted to the left of the red dashed line, while there are no photons produced to the right of the line. The dark shaded area marks the energy range, which is actually affected by the absorption, since at these times photons of these energies are produced in the source. In the light grey area the optical depth is irrelevant, because no more photons are produced at these energies. The behaviour described in the previous paragraph is obvious in figure \ref{fig:ssc}, but it also applies for the synchrotron target in figure \ref{fig:syn}. However, in figure \ref{fig:syn} the energy limits are higher than in figure \ref{fig:ssc}, and the highest emitted energies are captured by the synchrotron target with the given parameters.

In both figures \ref{fig:syn} and \ref{fig:ssc} the same parameters are used. One should note that these are extreme parameters with $\alpha=20$ and $l_{ec}=50$. For the synchrotron case this choice results in a high maximum optical depth of $\taum{syn} = 4080$. However, only the very highest energies of the SSC intensity are affected by the optical depth. At these energies the SSC intensity is already very low, as one can deduce from the very short emission time indicated by the red dashed line. Reducing $\alpha$ and $l_{ec}$ would result in a reduced $\taum{syn}$, but would not significantly change the absorbed energy region. Hence, the effect of the synchrotron target field on the resulting SED is minor for the given choice of parameters. Increasing, for example, the initial electron Lorentz factor would also result in a reduction of the maximum optical depth, but would on the other hand increase the energy region affected by photo-pair production, since the lower energy limit decreases and the SSC intensity reaches higher energies.

For figure \ref{fig:ssc} the extreme parameters are necessary to obtain some absorption at all. The parameters $\alpha=20$ and $l_{ec}=50$ correspond to a Compton dominance (ratio of the maximum value of the high energetic component to the maximum value of the synchrotron component) of roughly a factor $10^4$ \citep{zs12b}. Hence, the flux of the inverse Compton component is a factor $10^4$ higher than the flux in the synchrotron component. Such large values of the Compton dominance are not observed, and are usually $<100$.
%
%
\section{Photo-pair production for external Compton photon targets} \label{sec:ec}
As in the previous section, a general representation for the time dependence of the EC intensity can be found. Hence,
\begin{align}
I_{ec}(\eps,x) = I_{ec,0} \eps \left( v+px \right)^n G\left( q_i(\eps,x) \right) \HF{\eps_{ec,max}(x)-\eps} \label{eq:intecgen1} \eqc
\end{align}
with
\begin{align}
G(q_i) &= 2q_i\ln{q_i} + (1+2q_i)(1-q_i) + 2\epec \eps q_i (1-q_i) \label{eq:Gqi1} \eqc \\
q_i(\eps,x) &= \frac{\eps \left( v+px \right)^n}{\Gamec\gamma_0\left( 1- \frac{\eps}{\gamma_0} \left( v+px \right)^{n/2} \right)} \nonumber \\
&\approx \frac{\eps \left( v+px \right)^n}{\Gamec\gamma_0} \label{eq:qi1} \eqc \\
\eps_{ec,max}(x) &= \frac{\Gamec\gamma_0}{\left( v+px \right)^n \left( 1+\frac{\Gamec}{\left( v+px \right)^{n/2}} \right)} \label{eq:epsecmax1} \eqc \\
\Gamec &= 4\epec\gamma_0 \label{eq:gamec1} \eqd
\end{align}
The approximation in equation (\ref{eq:qi1}) holds, because the Heaviside function in equation (\ref{eq:intecgen1}) sets the intensity to zero before the bracket in the denominator of $q_i$ can deviate significantly from unity. $\Gamec$ is the Klein-Nishina parameter for EC scattering.

Inserting equation (\ref{eq:intecgen1}) into equation (\ref{eq:tauggint1}) the optical depth can be easily computed:
\begin{align}
\taug(\epss,x) = & \tau_{ec} \epss^{-1} \left( v+px \right)^n G\left( q_i(\epss,x) \right) \nonumber \\
& \times \HF{\epss - \frac{2}{\eps_{ec,max}(x)}} \label{eq:tauec1} \eqd
\end{align}
Here,
\begin{align}
\tau_{ec} &= 9.3\cdot 10^{-23} \frac{\alpha^2 \lec l_{ec} R_{15} b^2}{\gamf^4 \epec^2} \label{eq:tau0ec1} \eqc \\
q_i(\epss,x) &= \frac{2 \left( v+px \right)^n}{\epss\Gamec\gamma_0} \label{eq:qi2} \eqd
\end{align}
From the value of $\tau_{ec}$ it is obvious that a large optical depth requires rather extreme parameter settings, although one should keep in mind that generally $\epec\ll 1$.

In order to obtain some meaningful results, the function $G(q_i)$ must be approximated, since in the current form the equation $\tau_{ec}(\epstra{ec},x)=1$ cannot be solved. However, for a rather large parameter range $G(q_i)\approx 1$, and therefore this approximation is used to get a rough estimate of the transition energy. Now, the inversion is quite simple, and the transition energy becomes
\begin{align}
\epstra{ec} = \tau_{ec} \left( v+px \right)^n \label{eq:epstraec1} \eqd
\end{align}

\begin{figure}
	\centering
		\includegraphics[width=0.48\textwidth]{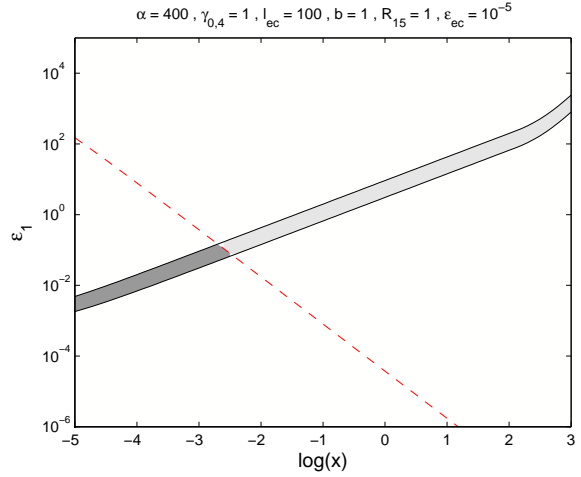}
	\caption{The plot displays the photon energy $\epss$ versus time. The shaded area shows the optically thick energy range due to interactions with an EC target field. The darker area shows the range where the optical depth actually applies, since the red dashed curve shows the cut-off energy of the SSC intensity at the respective time. Parameters are given at the top.}
	\label{fig:ec}
\end{figure}

The lower bound in figure \ref{fig:ec} is the application of the Heaviside function in equation (\ref{eq:tauec1}). The optically thick band is rather narrow despite the large value of $\alpha=400$. Therefore, the earlier statement that extreme parameter settings might be necessary has been confirmed. 

The dark shaded area in figure \ref{fig:ec} is the optically thick region for SSC photons, used again as an example for the high energy intensity. However, from the extreme parameter settings one can conclude that $\gamma-\gamma$ absorption with an EC photon target is not an issue for blazars.
%
%
\section{Photo-pair production for external photon targets} \label{sec:ext}
In this section the photo-pair production by an external photon target is calculated. It is kept on a basic level, since the absorption by such a photon field is discussed extensively in the literature, for example \cite{dp03,r07,ps10,sp11,sp14,dea09,dea14}, and many more. One can think of a lot of different external photon targets, such as the accretion disk photons, the broad line region (BLR) or the dusty torus. These possibilities differ in size and energy content, and thus their contribution to the optical depth around the emission blob strongly depends on the distance of the blob to the central black hole. A strong debate continues in the literature about the location of the ``blazar emission zone'', as can also be seen in the above given citations.

As mentioned in section \ref{sec:gamgamabs} the high energy photons produced in the emission blob can be absorbed in the entire external photon region. Therefore the path length becomes $R=R_{ex}$, which is scaled below as $R_{ex} = R_{pc}$pc, because $1$pc is a typical upper limit of the BLR size.

In \cite{zs12b} the external photon target for the inverse Compton scattering was modelled as a line-like $\delta$-function (see equation (\ref{eq:defuecline})), which could be produced by the BLR. Since, however, the integral of the optical depth already contains a delta-function, the delta function of the external photons is exchanged by a Gaussian in order to keep a similar model:
\begin{align}
\DF{\eps-\epec} \rightarrow \frac{1}{\sqrt{\pi}\sigma} e^{-\frac{\left( \eps-\epec \right)^2}{\sigma^2}} \label{eq:gaussian} \eqc
\end{align}
where $\sigma$ marks the widths of the Gaussian. It is a free parameter, which strongly influences the resulting energy range of the optical depth, as will be discussed later.

Since the number of photons and the energy density are connected by the relation $u(\eps) = mc^2\eps n(\eps)$, the photon number density becomes
\begin{align}
n(\eps) = \frac{4\Gamma_b^2u_{ec}^{\prime}}{3\sqrt{\pi}mc^2} \frac{1}{\sigma\eps}e^{-\frac{\left( \eps-\epec \right)^2}{\sigma^2}} \label{eq:next} \eqd
\end{align}
From the definition of $l_{ec}$, one can rewrite the energy density in external photons as $u_{ec}^{\prime} = 3l_{ec}u_B / 4\Gamma_b^2$. Then, the optical depth becomes
\begin{align}
\taug(\epss) = \frac{\tau_{ext}}{\sigma} e^{-\frac{\left( \frac{2}{\epss} - \epec \right)^2}{\sigma^2}} \label{eq:tauext1} \eqc
\end{align}
with
\begin{align}
\tau_{ext} = 1.9\cdot 10^{-2} R_{pc}l_{ec}b^2 \eqd \label{eq:tau0ec}
\end{align}

\begin{figure}
	\centering
		\includegraphics[width=0.48\textwidth]{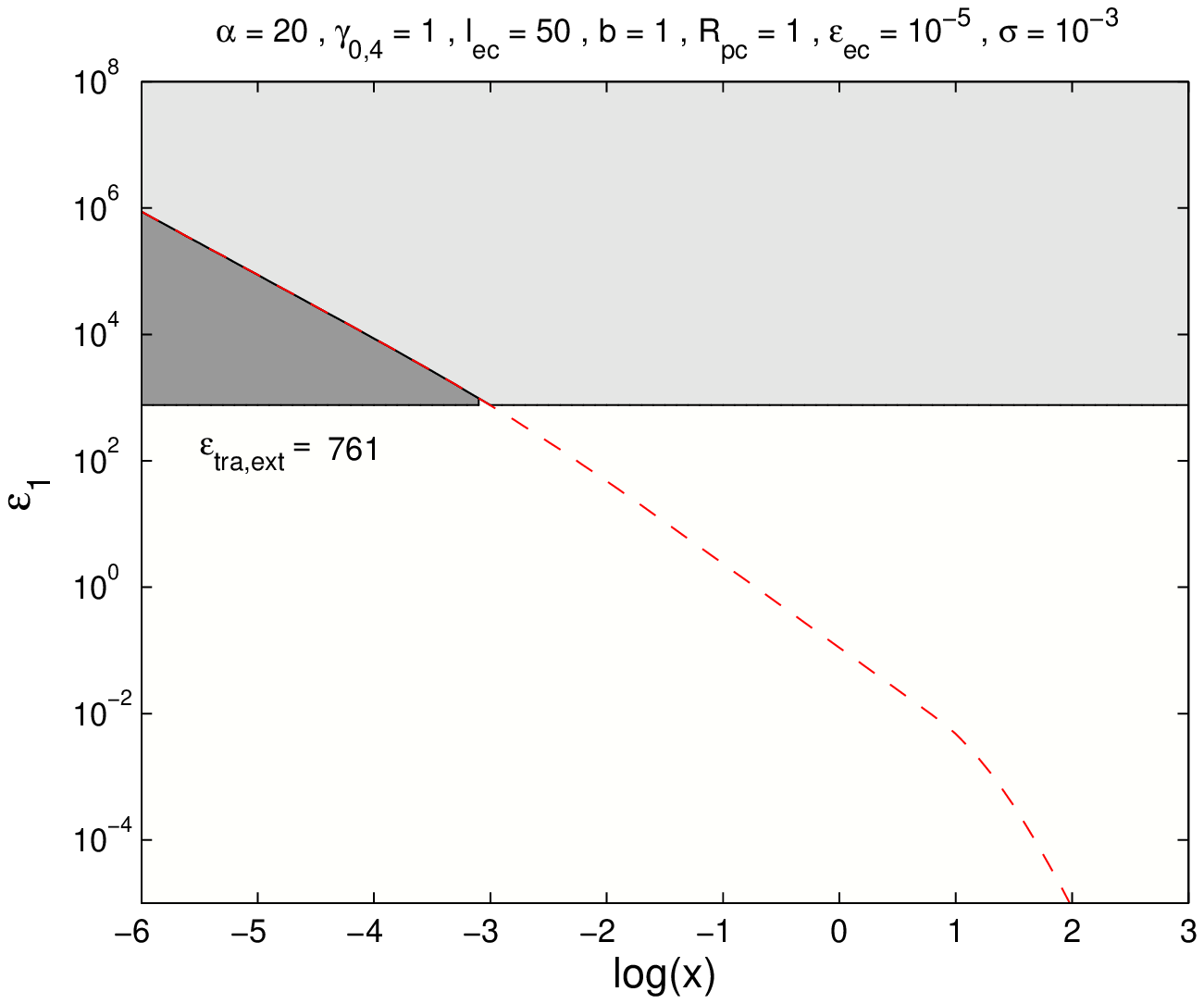}
	\caption{The plot displays the photon energy $\epss$ versus time. The shaded area shows the optically thick energy range due to interactions with a soft external photon target field. The darker area shows the range where the optical depth actually applies, since the red dashed curve shows the cut-off energy of the SSC intensity at the respective time. The value of $\epstra{ext}$ is indicated. Parameters are given at the top ($\alpha$ and $\gamf$ are also given, since they are needed for the SSC intensity cut-off).}
	\label{fig:ext}
\end{figure}

The consequence of equation (\ref{eq:tauext1}) is twofold. Firstly, the optical depth is obviously not time-dependent. Secondly, there is no distinct maximum. This implies that one can search for the energy $\epstra{ext}$ where $\taug(\epstra{ext})=1$, and for all energies above the source is optically thick for all times. The transition energy can be easily computed:
\begin{align}
\epstra{ext} = \frac{2}{\epec + \sigma \sqrt{\ln{\left( \frac{\tau_{ext}}{\sigma} \right)}}} \label{eq:epstraext1} \eqd
\end{align}
For $\sigma\rightarrow 0$ (i.e., the Gaussian becomes a $\delta$-function), the second summand in the denominator approaches zero and the transition energy becomes $\epstra{ext}=2/\epec$. However, in this case $\taug\rightarrow 0$, and therefore a fiducial value of $\sigma=10^{-3}$ is chosen for figure \ref{fig:ext}. As before, this figure shows the emitted energy $\epss$ plotted versus time, and the optically thick regime is marked in grey, while the red dashed line marks the exemplary SSC intensity cut-off.

%
%
%
\section{Influence on the total SED} \label{sec:sed}
\begin{figure*}
\begin{minipage}{0.49\linewidth}
\centering \resizebox{\hsize}{!}
{\includegraphics{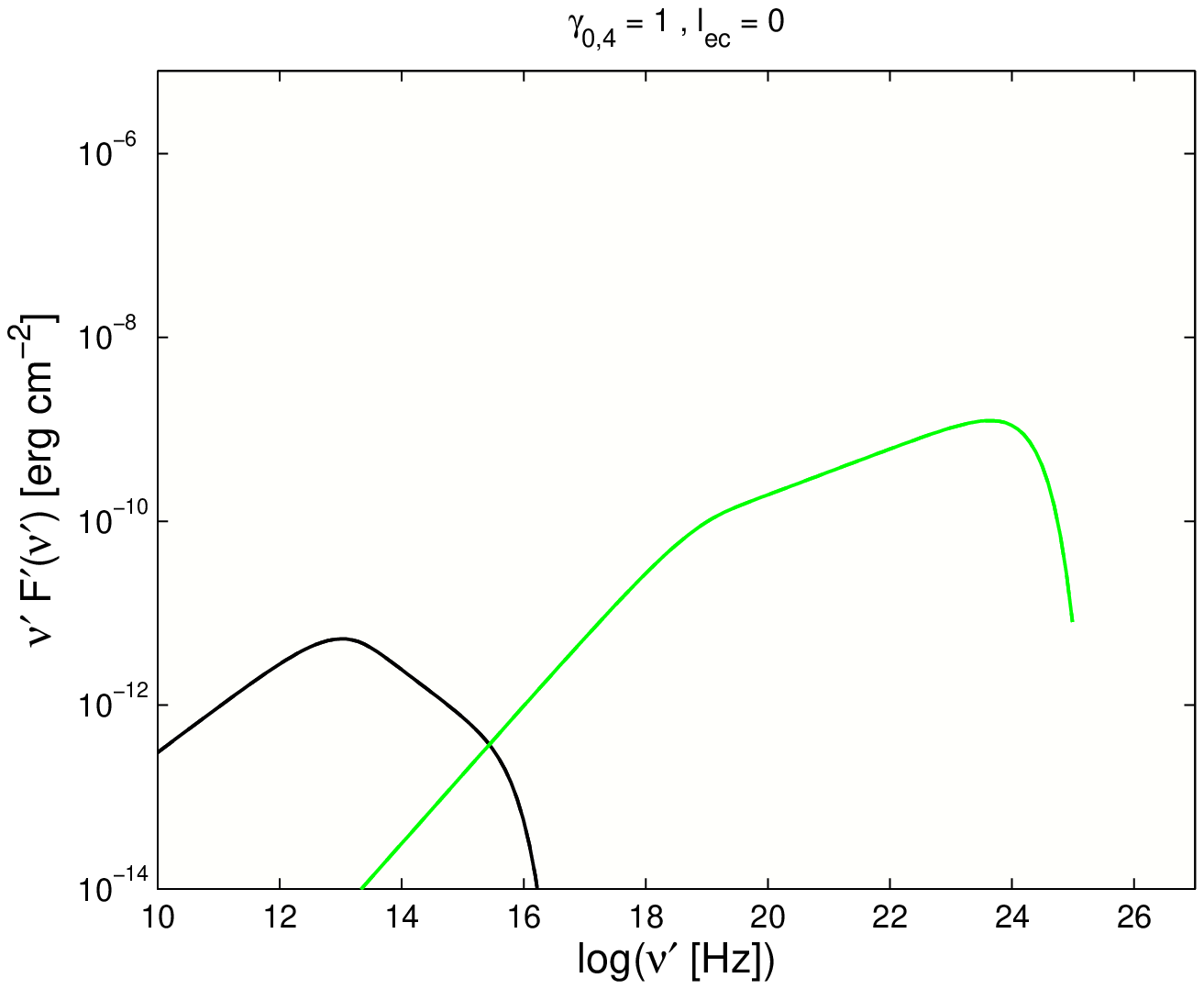}}
\end{minipage}
\hspace{\fill}
\begin{minipage}{0.49\linewidth}
\centering \resizebox{\hsize}{!}
{\includegraphics{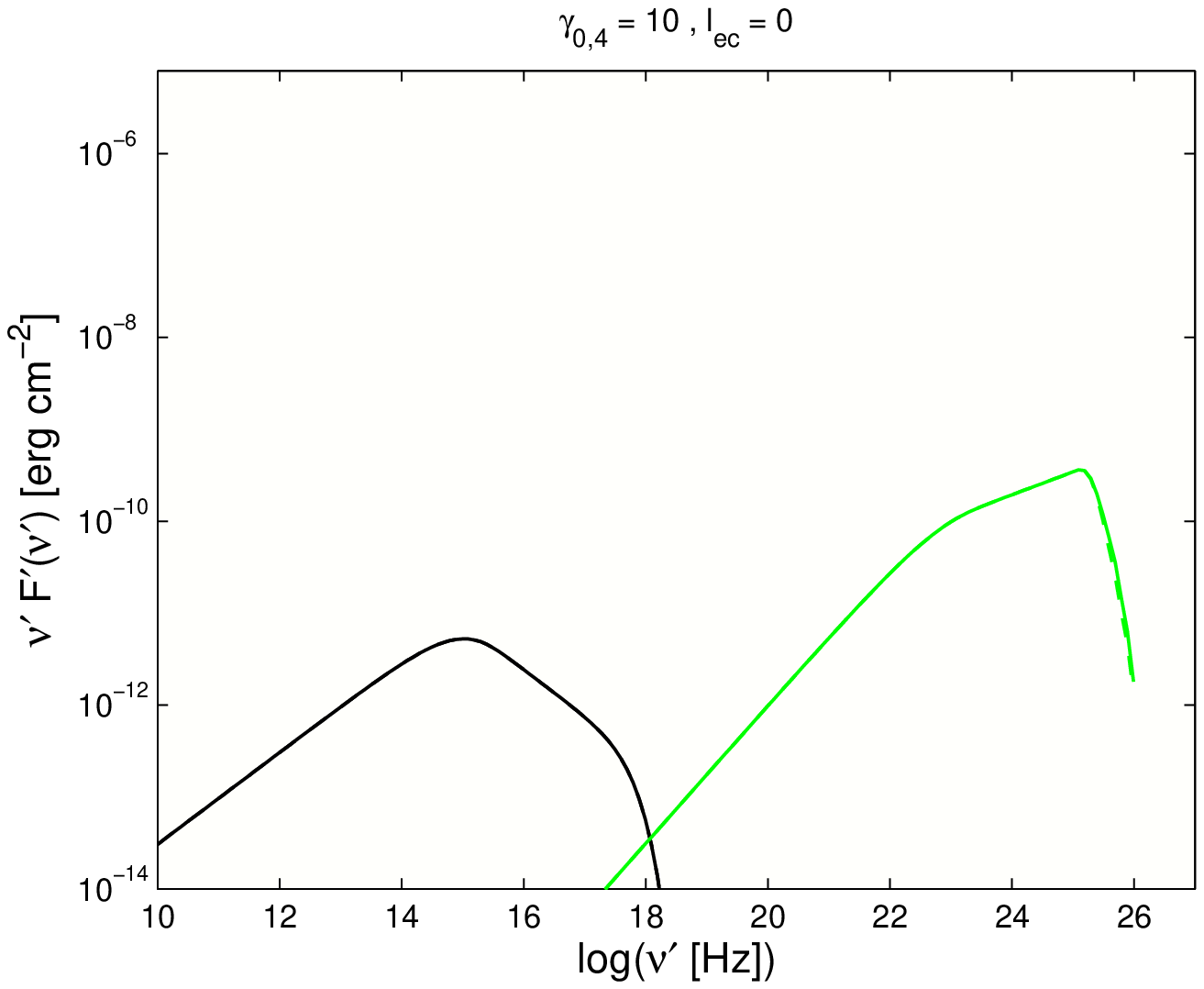}}
\end{minipage}
\newline
\begin{minipage}{0.49\linewidth}
\centering \resizebox{\hsize}{!}
{\includegraphics{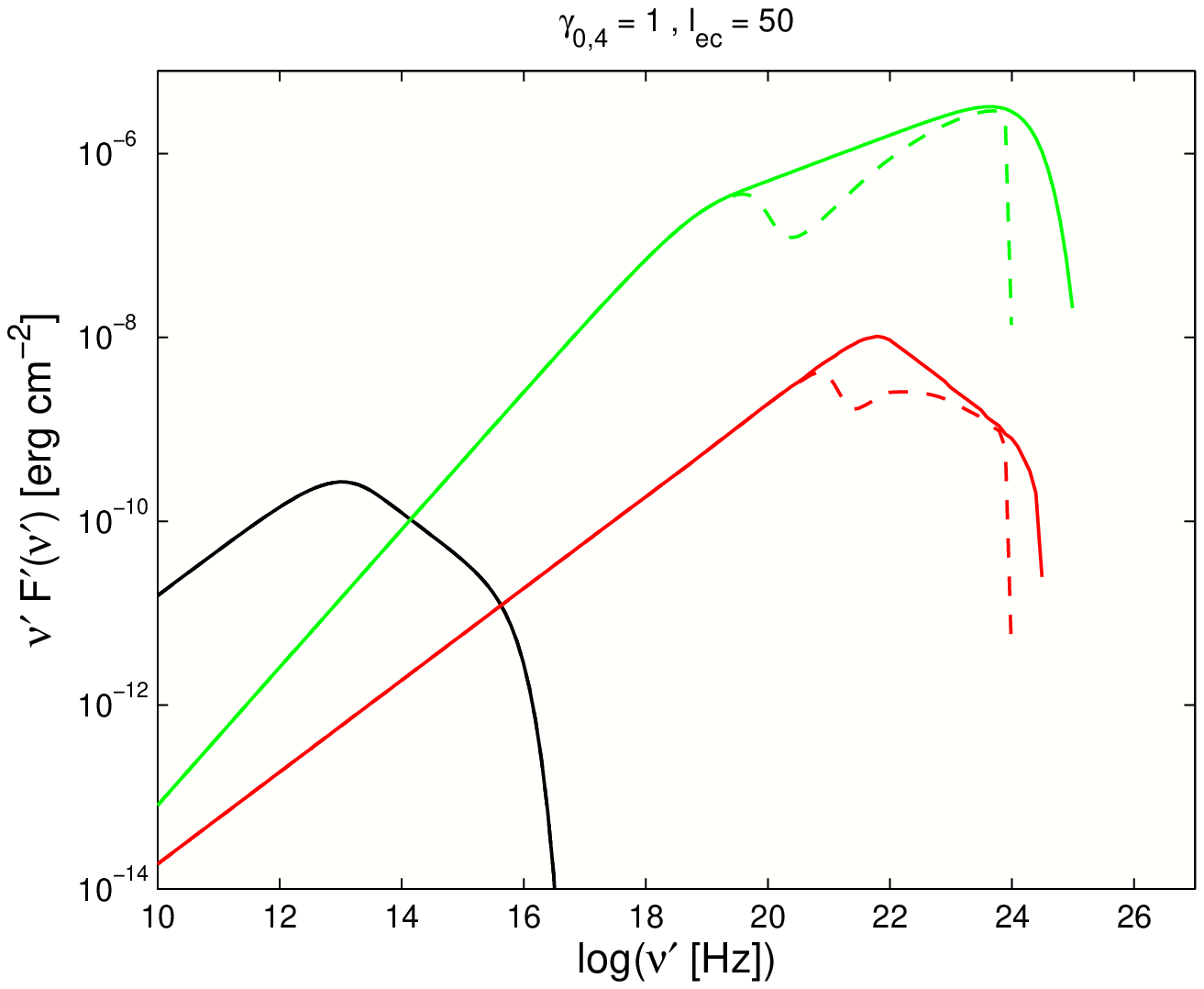}}
\end{minipage}
\hspace{\fill}
\begin{minipage}{0.49\linewidth}
\centering \resizebox{\hsize}{!}
{\includegraphics{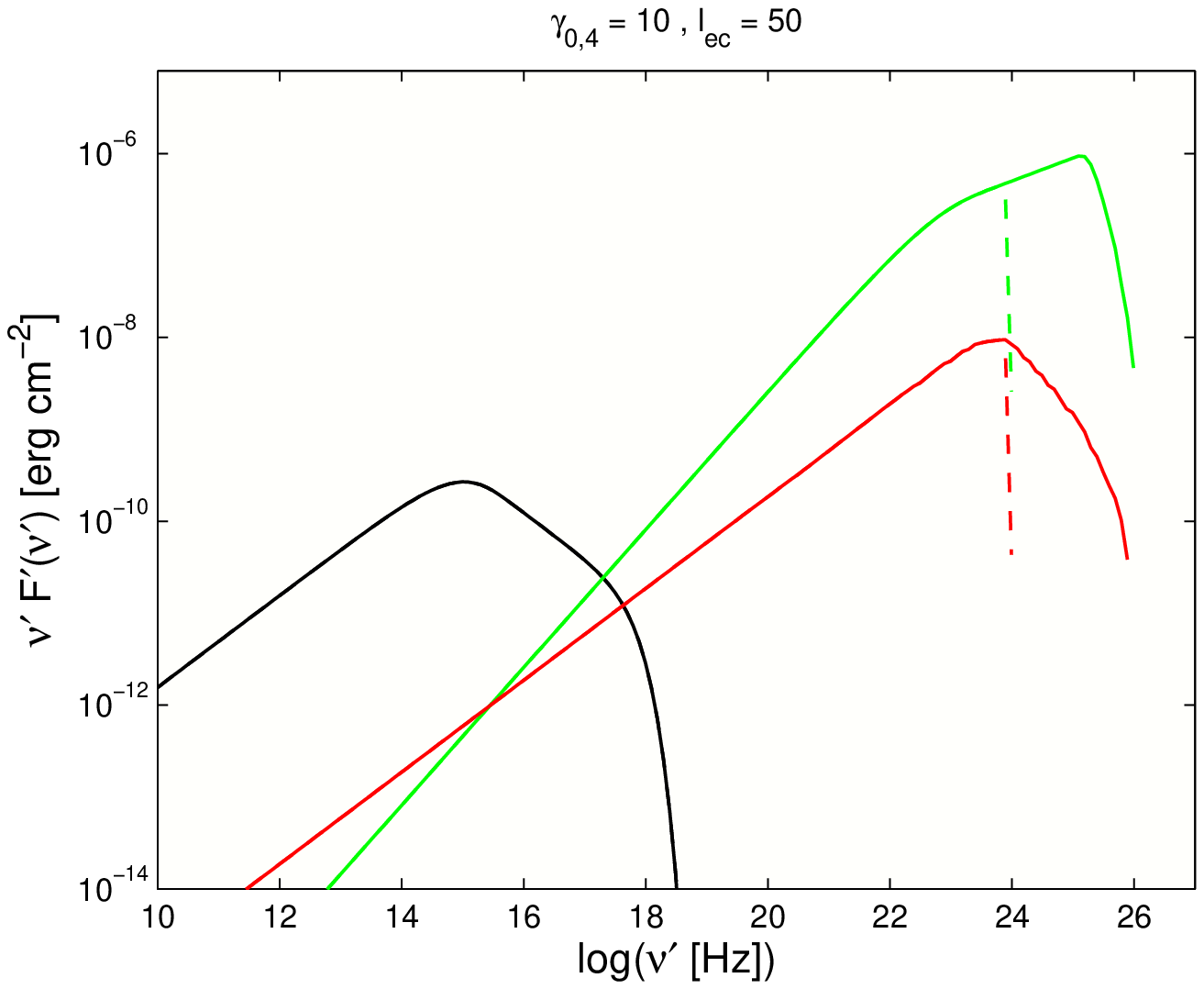}}
\end{minipage}
\caption{Total observed SEDs $\nu^{\prime} F^{\prime}(\nu^{\prime})$ versus observed frequency $\nu^{\prime} = \eps \delta_D mc^2/h(1+z)$. The solid lines are optically thin synchrotron (black), SSC (green) and EC (red) SEDs, while the dashed lines include absorption. Varied parameters are given at the top, while the constant parameters are given in table \ref{tab:params}.}
\label{fig:sed}
\end{figure*} 
Having calculated the optical depth $\taug$ for the various target photon fields, the optically thin intensities $I_{i}(\eps,x)$ of section \ref{sec:intdis} have to be modified as
\begin{align}
I_{\tau,i}(\eps,x) = I_{i}(\eps,x) \frac{1-e^{-\taug^{int}(\eps,x)}}{\taug^{int}(\eps,x)} e^{-\taug^{ext}(\eps)} \label{eq:optthickint} \eqc
\end{align}
where $\taug^{int}(\eps,x)$ is the sum of the internal optical depths, namely synchrotron, SSC, and EC. The optical depth due to the external photon field must be treated separately, and is denoted by $\taug^{ext}(\eps)$. The distinction between $\eps$ and $\epss$ has been dropped.

With equation (\ref{eq:optthickint}) the optically thick intensity distributions for all energies and times is known. However, observations are usually taken over a longer time scale, and therefore the intensities $I_{\tau,i}$ have to be integrated with respect to time as well, giving the SEDs $\eps F(\eps)$ after a formal multiplication with the photon energy $\eps$. Finite integration limits can be used to obtain fractional SEDs, which give valuable insights into the evolution of the SED. Here, only the total SEDs are calculated with integration from zero to infinity, i.e. over the entire flaring event. Transforming to the observer's frame (primed quantities) the total SEDs become
\begin{align}
\nu^{\prime} F_{i}^{\prime}(\nu^{\prime}) = \frac{R^2 \delta_D^4 (1+z)}{d_L^2} h\nu^{\prime} \intl_0^{\infty} I_{\tau,i} \left( \nu^{\prime}\frac{h(1+z)}{\delta_D mc^2},t  \right) \td{t} \label{eq:SEDdef} \eqc
\end{align}
where $\delta_D$ is the Doppler factor, $z$ the red shift, and $d_L$ the luminosity distance of the source. Although the proper transformations have been applied the absorption by the extragalactic background light has been neglected. It would cause red shift dependent absorption at the highest energies, similar to the absorption of the external soft photon field. 

Examples are plotted in figure \ref{fig:sed}. The solid lines indicate the optically thin SEDs, while the dashed lines show the SEDs with the influence of the photo-pair production. Varied in each plot are the scaled initial electron Lorentz factor $\gamf$ and the external Compton parameter $l_{ec}$. In the top row of figure \ref{fig:sed} $l_{ec}=0$ implies no soft external photon field and, of course, neither an EC component nor absorption effects by that field. The other parameters are kept constant (see table \ref{tab:params}) in all plots, most importantly $\alpha = 20$ resulting in a strong dominance of the SSC component.

As discussed in section \ref{sec:synssc} the effect of the synchrotron target is only minor. In the top left plot in figure \ref{fig:sed} it influences only the highest energies, which are not even visible in the plot, since they are beyond the exponential cut-off. The same is true for the lower left plot, while in the lower right plot the effect is probably hidden by the strong influence of the external photon field. In the upper right plot the influence is barely discernible, because of the strong impact by the Klein-Nishina reduction.

The SSC target field becomes important for more extreme settings in the lower left plot, where $l_{ec} = 50$. The effect on intermediate energies is clearly visible. The slight shift of the dip position between the SSC and EC component can be explained by the different evolutions of the respective intensities with respect to time. The SSC intensity evolves much faster than the EC intensity, which means that the maximum of the intensity distribution of the SSC moves faster to lower energies than the maximum of the EC in any given time interval.\footnote{This is also obvious from the position of the break in the SSC and EC component. Since the break is due to the change in the cooling behaviour at time $x=x_c$, the lower break energy of the SSC component implies a faster evolution.} And since the optical depths is also strongly time-dependent (at least for the synchrotron, SSC, and EC targets) the position of the dip in the total SED differs between the SSC and the EC component.

As mentioned in section \ref{sec:ec} the influence of the EC target can be safely neglected.

The lower row in figure \ref{fig:sed} includes the external photon field. Apart from the emerging EC component, the severe effect of the absorption by the external field is obvious. Irrespective of the initial internal parameters both SSC and EC component are cut-off at roughly $10^{24}$Hz in both plots. Hence, potential dips at higher energies caused by the internal fields are not visible.
\begin{table}
\begin{center}
	\begin{tabular}{l|c|l}
		\multicolumn{2}{c}{Parameter} & Value \\
		\hline
		Injection parameter & $\alpha$ & $20$ \\
		Magnetic field & $b$ & $1$ \\
		External photon energy & $\epec$ & $10^{-5}$ \\
		Blob radius & $R_{15}$ & $1$ \\
		BLR radius & $R_{pc}$ & $1$ \\
		Gaussian widths & $\sigma$ & $10^{-3}$ \\
		Doppler factor & $\delta_D$ & $10$ \\
		Red shift & $z$ & $0.1$ \\
		Luminosity distance & $d_L$ & $1.43\cdot 10^{27}$cm
	\end{tabular}
\caption{Constant parameters in figure \ref{fig:sed}.}
\label{tab:params}
\end{center}
\end{table}
\subsection{Minimum variability and the Doppler factor}
Due to retardation effects the chosen size of the emission region $R_{15}=1$ implies a minimum variability time scale of $t_{var}^{\prime}=R_b(1+z)/(c\delta)\approx 55$min in the observer's frame. Reducing the size of the emission region by 1 order of magnitude would make $t_{var}^{\prime}$ comparable to the fastest observed variabilities on the order of a few minutes \citep{c04,aeaH07,aeaM07}. According to the results presented above, the optical depth would also be reduced. However, the observed radiation power would be reduced like-wise, which is in contrast to the observational fact that short variability time scales correspond to extraordinary high flux states.

Hence, in order to verify if the time-dependent injection model is able to explain the most rapid variabilities, the string of arguments laid out by \cite{dg95} is followed. According to equation (\ref{eq:tauggint1}), the optical depths due to an internal photon field equals
\begin{align}
\taug(\epss,t) &= \frac{R_b\sigma_T}{3c} I\left( \eps=\frac{2}{\epss} , t \right) \nonumber \\
&= \frac{d_L^2\sigma_T}{3cR_b\delta^3(1+z)} I^{\prime}\left( \eps^{\prime} = \frac{2\delta^2}{\epss^{\prime}(1+z)^2}, t^{\prime} \right) \label{eq:timedisc1} \eqd
\end{align}
In the second line the transformation to the observer's frame has been applied. Note that the intensity transformation only requires three powers of the Doppler factor, while the transformation of the time integrated fluence requires four powers. This follows from the invariance of both $I(\epsilon,t)/\epsilon^3$ and $\td{t}/\epsilon$ \citep{dm09}.

Using the requirement that $R_b\leq c\delta t_{var}^{\prime}/(1+z)$, the minimum optical depth becomes
\begin{align}
\taug(\epss,t) >& \tau_{\gamma\gamma,min}(\epss,t) \nonumber \\
&= \frac{d_L^2\sigma_T}{3c^2\delta^4} \frac{I^{\prime}\left( \eps^{\prime} = \frac{2\delta^2}{\epss^{\prime}(1+z)^2}, t^{\prime} = t_{var}^{\prime} \right)}{t_{var}^{\prime}} \label{eq:timedisc2} \eqd
\end{align}
There are two reasons to set the time variable in the intensity equal to the variability time scale. Firstly, the intensity value at the variability time scale $t_{var}^{\prime}$ is the maximum of the lightcurve for the interesting energies, since only the highest energies of each component exhibit variability time scales as short as the light crossing time scale \citep{zs13,z14}. As a matter of fact, the cited minute-short variabilities are observed at the highest energies of either the synchrotron or the inverse Compton component. Secondly, the intensity distribution in the source at $t_{var}^{\prime}$ is a good representative of the state of the source for the entire time period since the beginning of the flare. This is especially true, if the internal retardation is taken into account. Even though the electrons have considerably cooled and the emitted photons at $t^{\prime}=t_{var}^{\prime}$ are already less energetic than those emitted at the beginning of the flare, the source is still filled with the higher energetic photons from earlier times, since most of them did not have had the time to leave the source and are still available for absorption processes. This has been neglected in the preceding discussion for reasons of simplicity. The effect on the cooling process by these retarded photons will be discussed elsewhere.

In an optically thin source $\taug(\epss,t)<1$, and the resulting inequality can be solved for the Doppler factor. Hence,
\begin{align}
\delta >& \delta_{min} \nonumber \\
&= \left( \frac{d_L^2\sigma_T}{3c^2} \frac{I^{\prime}\left( \eps^{\prime} = \frac{2\delta^2}{\epss^{\prime}(1+z)^2}, t^{\prime} = t_{var}^{\prime} \right)}{t_{var}^{\prime}} \right)^{1/4} \nonumber \\
&\approx 150 \left( \frac{I^{\prime}\left( \eps^{\prime} = \frac{2\delta^2}{\epss^{\prime}(1+z)^2}, t^{\prime} = t_{var}^{\prime} \right)}{t_{var}^{\prime}} \right)^{1/4} \eqd \label{eq:timedisc3}
\end{align}
This is the same result as in \cite{dg95}. Thus, the time-dependent injection model cannot resolve the strong requirement on the Doppler factor by the typical one-zone model, which is in conflict with VLBI observations where the observed Lorentz factors are on the order of a few (the so-called ``Doppler factor crisis'', e.g. \citet{hs06,bfr08,fdb08}, and many more). 

Nevertheless, due to the severe consequences of the time-dependent injection on the spectrum and the light curves, as shown in previous papers, a wide utilization is encouraged. A combination with models that try to explain the rapid variability, such as jets-in-a-jet \citep{gub09,bg12}, magneto-centrifugal acceleration \citep{gtbc09}, jet-star interactions \citep{bea12}, and others, might have interesting consequences justifying further studies.

%
%
\section{Summary and conclusions} \label{sec:discon}
In this paper the time-dependent character of the optical depth $\taug$ in blazars subject to a time-dependent injection scenario is discussed. For high energy photons the photo-pair production optical depth is calculated using the simplifying delta-function approach for the cross section. In all cases the potentially time-dependent transition energies $\eps_{tra}(x)$ between the optically thin and thick regimes are calculated satisfying the relation $\taug(\eps_{tra},x) = 1$. 

The photo-pair production of high energy photons with a target photon field is most relevant for steady photon targets. A line-like distribution of external photons is given in this work as an example. Above a certain energy threshold the source is optically thick, just as expected, with an increasing value of $\taug$ for increasing values $\epss$ of the high energy photon energies.

Apart from the external photons other possible target photon fields are abundant in the source, namely the synchrotron, the SSC and the EC photons produced by the relativistic electrons of the emission blob. Since these target photon fields are time-dependent, some at first sight strange results emerge for the optical depth. Namely, there are two solutions for the equation $\taug(\eps_{tra},x) = 1$, resulting in time-dependent upper and lower energy limits for the optically thick regime. For synchrotron photons these limits are so high that only the highest energies are affected. On the other hand, these are usually beyond the cut-off energy of the high energetic component, and thus only have a minor impact.

If the limits for the optically thick regime due to an SSC target exist, they are at first at rather low energies. At later times these limits increase to higher energies. However, since the time dependence of the respective intensities imply a decreasing cut-off time for increasing energies, no more high energetic photons are produced when the source becomes optically thick for them. Hence, high energy photons might leave the source while medium or low energetic photons cannot. This would cause a dip in the high energy component in the MeV energy regime -- an energy window currently not covered by any operational telescope. This might be an interesting target for the planned \textsc{AstroMeV}\footnote{http://astromev.in2p3.fr/} and \textsc{Gamma-Light} \citep{mea13} missions. On the other hand, a non-detection of such a feature would put further constraints on the source parameters.

The EC target requires extreme parameter values in order to cause some absorption. Hence, absorption by EC photons can be safely ignored.

To summarize, the optical depths due to the time-dependent internal photon fields is not a big issue. For purely linearly cooling sources (i.e. $\alpha<1$) the optical depth due to photo-pair production can be neglected. Even for $\alpha>1$ only the SSC photons might have a visible impact on the SED, as can be seen in the example SEDs in figure \ref{fig:sed}. However, even for the SSC target a high value for the parameter $l_{ec}$ is necessary implying an unrealistically large Compton dominance (c.f. the discussion at the end of section \ref{sec:synssc}). 

With respect to the so-called ``Doppler factor crisis'' the simple change in the one-zone model from continuous to time-dependent injection does not solve the problem. It might, however, be useful in combination with other models.

For the sake of completeness, the calculation of the synchrotron-self absorption optical depths is presented in appendix \ref{app:ssa}. This applies for low synchrotron photon energies, and is observed in some blazars as a steeper spectrum at radio wavelengths compared to an optically thin spectrum. In such cases the synchrotron-self absorption turn-over energy could be used to constrain the parameters. The time-dependent injection model challenges this simple deduction, since the turn-over energy is not a constant any more but becomes strongly time dependent. From an initial low value the turn-over energy increases to a maximum and then decreases until the source becomes totally optically thin at low synchrotron energies. For a well sampled spectrum it might be possible to deduce the maximum energy. Since it depends only on three free parameters, it would certainly help to constrain the free parameters. Otherwise, precise knowledge of the time of observation with respect to the onset of the flare (i.e., time of injection) is necessary in order to make any useful statement.

In conclusion, the optical depth can potentially be used to constrain the free parameters, but is not a limitation to the time-dependent injection model itself. 
%
%
\section*{Acknowledgement}
The author is grateful to Markus B\"ottcher for helpful discussions, as well as to the anonymous referee for detailed and constructive reports, which helped significantly improving the manuscript. \\
This work was supported by the German Ministry for Education and Research (BMBF) through Verbundforschung Astroteilchenphysik grant $05A11VH1$.
%
%
\appendix
%
\section{Calculation of the intensities} \label{app:int}
Here, a short summary of the calculation of the intensities from section \ref{sec:intdis} is presented. First, the electron distribution function is obtained from the kinetic equation, which is used afterwards to calculate the emerging optically thin photon intensities.
%
\subsection{Solution of the kinetic equation} \label{app:intkin}
Following \cite{k62}, the electron distribution function $n(\gamma,t)$ with the electron Lorentz factor $\gamma$ and time $t$ can be calculated from the kinetic equation:
\begin{align}
\frac{\pd{n(\gamma,t)}}{\pd{t}} - \frac{\pd{}}{\pd{\gamma}} \left[ |\dot{\gamma}|_{tot} n(\gamma,t) \right] = Q(\gamma,t) \eqd \label{eq:kineq1}
\end{align}
The total cooling term $|\dot{\gamma}|_{tot}$ contains three terms, namely the linear synchrotron and EC cooling terms, and the non-linear, time-dependent SSC cooling term. Hence,
\begin{align}
|\dot{\gamma}|_{tot} &= |\dot{\gamma}|_{syn} + |\dot{\gamma}|_{ec} + |\dot{\gamma}|_{ssc} \nonumber \\
&= D_0\lec \gamma^2 + A_0 \gamma^2 \intl_0^{\infty} \gamma^{\prime 2} n(\gamma^{\prime},t) \td{\gamma^{\prime}} \label{eq:totcool} \eqc
\end{align}
where $l_{ec}$ combines both linear cooling terms according to equation (\ref{eq:deflec}). The lower limit in the integral is an approximation for large Lorentz factors.

A simple form of the injection $Q(\gamma,t)$ is chosen:
\begin{align}
Q(\gamma,t) = Q_0 \DF{\gamma-\gamma_0} \DF{t} \label{eq:parinj} \eqc
\end{align}
that is a single burst of particles with number density $Q_0$, and Lorentz factor $\gamma_0$ at time $t=0$. The delta-function in time might seem as a very simple form for a particle injection. It has, however, the important consequence that the particles cannot reach equilibrium, which results in the non-linearity of the SSC cooling term. In equilibrium conditions (i.e., constant injection in time) the non-linearity is inhibited, because the integral in equation (\ref{eq:totcool}) yields a constant.

The derivation of the solution to equation (\ref{eq:kineq1}) under the given scenario has been computed by \cite{sbm10}. It turns out that the initial conditions, summarized in the injection parameter $\alpha$ (c.f. equation (\ref{eq:defalpha})), define the result. For $\alpha<1$ the electron distribution function becomes
\begin{align}
n(\gamma,x) = Q_0 \DF{\gamma - \frac{\gamma_0}{1+x}} \HF{\gamma_0-\gamma} \label{eq:eldis0} \eqc
\end{align}
which is a purely linear solution. For $\alpha>1$ the result is twofold:
\begin{align}
n(\gamma,x<x_c) &= Q_0 \DF{\gamma - \frac{\gamma_0}{\left( 1+3\alpha^2 x \right)^{1/3}}} \HF{\gamma_0-\gamma} \label{eq:eldis1} \eqc \\
n(\gamma,x>x_c) &= Q_0 \DF{\gamma - \frac{\gamma_0}{\alpha_g+x}} \HF{\frac{\gamma_0}{\alpha}-\gamma} \label{eq:eldis2} \eqc
\end{align}
which is a non-linear solution for early times, and a modified linear solution for late times. 
%
\subsection{The synchrotron intensity} \label{app:intsyn}
The optically thin synchrotron intensity is calculated by
\begin{align}
I_{syn}(\eps,x) = \frac{R}{4\pi} \intl_{0}^{\infty} n(\gamma,x) P_{syn}(\eps,\gamma) \td{\gamma} \label{eq:appintsyn1} \eqc
\end{align}
where the photon energy has been normalized to the electron rest mass energy $\eps = E_{ph}/mc^2$. The synchrotron emission power of a single electron is
\begin{align}
P_{syn}(\eps,\gamma) = \frac{P_0mc^2\eps}{\gamma^2} CS\left( \frac{2mc^2\eps}{3E_0\gamma^2} \right) \label{eq:apppsyn1} \eqc
\end{align}
with $P_0 = 2\cdot 10^{24}$ erg$^{-1}$s$^{-1}$, and $E_0 = 1.856 \cdot 10^{20}b$ erg. The function $CS(w)$ can be approximated as
\begin{align}
CS(w) \approx a_0 w^{-2/3} e^{-w} \label{eq:defCS} \eqc
\end{align}
with $a_0=1.151275$ (\citet{cs86,cs88}).

Inserting the respective electron distribution functions into equation (\ref{eq:appintsyn1}) gives the intensities (\ref{eq:intsyn0}), (\ref{eq:intsyn1}), and (\ref{eq:intsyn2}). 
%
\subsection{The SSC intensity} \label{app:intssc}
The optically thin SSC intensity is calculated similarly to the synchrotron intensity, by replacing $P_{syn}$ with $P_{ssc}$ in equation (\ref{eq:appintsyn1}). The SSC emission power of a single electron is given by 
\begin{align}
P_{ssc}(\eps_s,\gamma) = & R mc^2\eps_s \intl_{0}^{\infty} \td{\eps} \intl_{\gamma_{min}}^{\infty} \td{\gamma} \frac{1}{\eps} n(\gamma,x) \nonumber \\
&\times P_{syn}(\eps,\gamma) \sigma_{KN}(\eps_s,\eps,\gamma) \label{eq:pssc1} \eqd
\end{align}
Here, $\eps_s$ is the normalized photon energy after scattering,\footnote{The subscript $s$ is dropped in the main body of the text, since the distinction is not necessary there.} and
\begin{align}
\gamma_{min} = \frac{\eps_s}{2} \left( 1+\sqrt{1+\frac{1}{\eps\eps_s}} \right) \eqd
\end{align}
The Klein-Nishina cross section is defined as
\begin{align}
\sigma_{KN}(\eps_s,\eps,\gamma) = \frac{3\sigma_T}{4\eps\gamma^2} G(q(\eps_s,\eps,\gamma)) \label{eq:defsigkn} \eqc
\end{align}
with the Thomson cross section $\sigma_T = 6.65\cdot 10^{-25}$ cm$^2$, and 
\begin{align}
G(q) =& 2q\ln{q} + (1+2q)(1-q) \nonumber \\
&+ 2\eps\eps_s q(1-q) \label{eq:defG} \eqc \\
q(\eps_s,\eps,\gamma) =& \frac{\eps_s}{4\eps\gamma (\gamma-\eps_s)} \label{eq:defq} \eqd
\end{align}

The resulting equation for the intensity contains three integrals, from which two can be easily solved due to the delta-functions in the electron distribution. The third integral was approximately solved by \cite{s09} using $q$ instead of $\eps$ as integration variable. The integral is of the form
\begin{align}
J(A) &= \intl_0^1 \frac{1}{q} CS\left( \frac{A}{q} \right) G(q) \td{q} \nonumber \\
&\approx \frac{63a_0}{100} A^{-2/3} e^{-A} \label{eq:intJapprox} \eqd
\end{align}

Inserting all these definitions, the intensities (\ref{eq:intssc0}), (\ref{eq:intssc1}), and (\ref{eq:intssc2}) emerge. 
%
\subsection{The EC intensity} \label{app:intec}
In \cite{zs12b} the intensity of an isotropic electron distribution scattering an angle-averaged external photon distribution has been derived as
\begin{align}
I_{ec}(\eps_s,x) = \frac{Rc\eps_s}{4\pi mc^2} \intl_0^{\infty}\td{\eps} \frac{u(\eps)}{\eps} \intl_0^{\infty} \td{\gamma} n(\gamma,x) \sigma_{KN}(\eps_s,\eps,\gamma) \label{eq:appintecgen1} \eqd
\end{align}
Here, $u(\eps)$ is the energy density of the external photon source in the blob frame, and $\sigma_{KN}(\eps_s,\eps,\gamma)$ is the Klein-Nishina cross section as given in the previous section.

For simplicity, \cite{zs12b} assumed a line-like distribution of external photons, which they modelled as
\begin{align}
u(\eps) = \frac{4\Gamma_b^2}{3} u_{ec}^{\prime} \DF{\eps-\epec} \label{eq:appextphotendens} \eqd
\end{align}
The fraction contains the angle-averaged transformation rule from the galactic system (where the external energy density $u_{ec}^{\prime}$ is defined) to the blob frame. The blob moves with a Lorentz factor $\Gamma_b$. The normalized energy of the line radiation is given as $\epec$, which is usually much smaller than unity.

Inserting the external photon density and the electron distribution into equation (\ref{eq:appintecgen1}) yields the respective intensities given in equations (\ref{eq:intec0}), (\ref{eq:intec1}), and (\ref{eq:intec2}). 
%
%
\section{SSC cut-off energy} \label{app:ssccoe}
Since the intensities from section \ref{sec:intdis} are time-dependent, the number of emitted photons of a specific energy $\eps$ depends on time, as well. According to the lightcurves calculated by \cite{zs13} and \cite{z14}, the cut-off time of the lightcurve increases for decreasing photon energy. This has interesting consequences for the optical depths as is discussed in the respective sections. 

In order to compare the time-dependent optical depth with the time-dependent high energy intensities, the SSC cut-off energy is shown in figures \ref{fig:syn} - \ref{fig:ext}. Following \cite{z14}, the cut-off energy is obtained as follows. Rewriting the argument of the exponential of the general intensity (\ref{eq:Issgen})
\begin{align}
\frac{\eps}{\epst}\left( v+px \right)^n &= \frac{\eps}{\epst} + \frac{\eps}{\epst} \left[ \left( v+px \right)^n -1 \right] \nonumber \\
&= \frac{\eps}{\epst} + A(\eps,x) \eqc
\end{align}
the cut-off energy is determined from the equation $A(\eps_{off},x)=1$, resulting in
\begin{align}
\eps_{off}(x) = \frac{\epst}{\left( v+px \right)^n -1} \label{eq:ssccoe} \eqd
\end{align}
This cut-off energy is shown in figures \ref{fig:syn} - \ref{fig:ext} as the red-dashed line.
%
%
\section{Synchrotron-self absorption} \label{app:ssa}
Synchrotron-self absorption is an important process at very low synchrotron energies. It cannot be treated as the previous cases. The following calculations have already been performed by \cite{z13phd}, and it is presented here in order to give a complete picture of the internal absorption processes. The relevant equations are well known (e.g. \citet[ch. 7.8]{dm09}). Here, only the calculation of the transition energy is presented. The details with respect to the spectrum and its time-dependent evolution can be found in \cite[ch. 3]{z13phd}.

The optical depths of synchrotron-self absorption can be written as
\begin{align}
\taus(\eps,x) = \frac{Rh^3}{8\pi m^3c^4\eps^2} \intl_0^{\infty} \frac{n(\gamma,x)}{\gamma^2} \frac{\td{}}{\td{\gamma}} \left[\gamma^2P_{syn}(\eps,\gamma)\right] \td{\gamma} \label{eq:ssagen1} \eqd
\end{align}
Here, $h=6.6261\cdot 10^{-27}$erg$\,$s is Planck's constant, and the spectral synchrotron power is given by
\begin{align}
P_{syn}(\eps,\gamma) = \frac{P_0mc^2\eps}{\gamma^2} CS\left( \frac{2mc^2\eps}{3E_0\gamma^2} \right) \label{eq:psyn1} \eqc
\end{align}
with
\begin{align}
CS(w) = a_0 w^{-2/3} e^{-w} \eqd
\end{align}
The specific values of the constant can be found in appendix \ref{app:intsyn}.

With these definitions, the derivative in equation (\ref{eq:ssagen1}) can be easily computed:
\begin{align}
\frac{\td{}}{\td{\gamma}}\left[ \gamma^{4/3} e^{-\frac{2mc^2\eps}{3E_0\gamma^2}} \right] = \frac{4}{3} \gamma^{1/3} e^{-\frac{2mc^2\eps}{3E_0\gamma^2}} \left[ 1+\frac{mc^2\eps}{E_0\gamma^2} \right] \label{eq:ssadev1} \eqd
\end{align}

Due to some complications stemming from equation (\ref{eq:ssadev1}) the optical depth cannot be evaluated in the general form as was done in sections \ref{sec:synssc} and \ref{sec:ec}. Instead, equation (\ref{eq:ssagen1}) has to be solved for each case of $\alpha$ individually.
%
\subsection{The case $\alpha<1$} \label{app:ssa01}
Inserting the electron distribution function for the case $\alpha<1$ [equation (\ref{eq:eldis0})] into equation (\ref{eq:ssagen1}), one immediately obtains
\begin{align}
\taus(\eps,x) =& I_{ssa} \left( \frac{\eps}{\epsn} \right)^{-5/3} \left( 1+x \right)^{5/3} \nonumber \\
&\times \left[ 1+\frac{3}{2}\frac{\eps}{\epsn}\left( 1+x \right)^2 \right] e^{-\frac{\eps}{\epsn}\left( 1+x \right)^2} \label{eq:taus01a} \eqc 
\end{align}
with $I_{ssa} = 1.53\cdot 10^{-13} \alpha^2 / b\gamf^7$. The relation $I_{ssa}<1$ results in $\alpha<2.5\cdot 10^6 \gamf^{7/2}b^{1/2}$. Although this relation should hold in basically all circumstances, it is not a strict requirement. However, it has some consequences that become obvious below.

Setting $\taus(\epstra{ssa},x) = 1$, the time-dependent behaviour of the transition energy $\epstra{ssa}$ can be obtained. The steps from section \ref{sec:synssc} can be used again in order to obtain approximative results. Approximating first for small values of the argument of the exponential $\eps(1+x)^2/\epsn\ll 1$, the transition energy becomes
\begin{align}
1 &\approx I_{ssa} \omega_{tra,01}^{-5/3} \left( 1+x \right)^{5/3} \nonumber \\
\Rightarrow \omega_{tra,01} &\approx I_{ssa}^{3/5} \left( 1+x \right) \label{eq:epstra01a} \eqc
\end{align}
where $\omega_{tra} = \epstra{ssa} / \epsn$.

\begin{figure}
	\centering
		\includegraphics[width=0.48\textwidth]{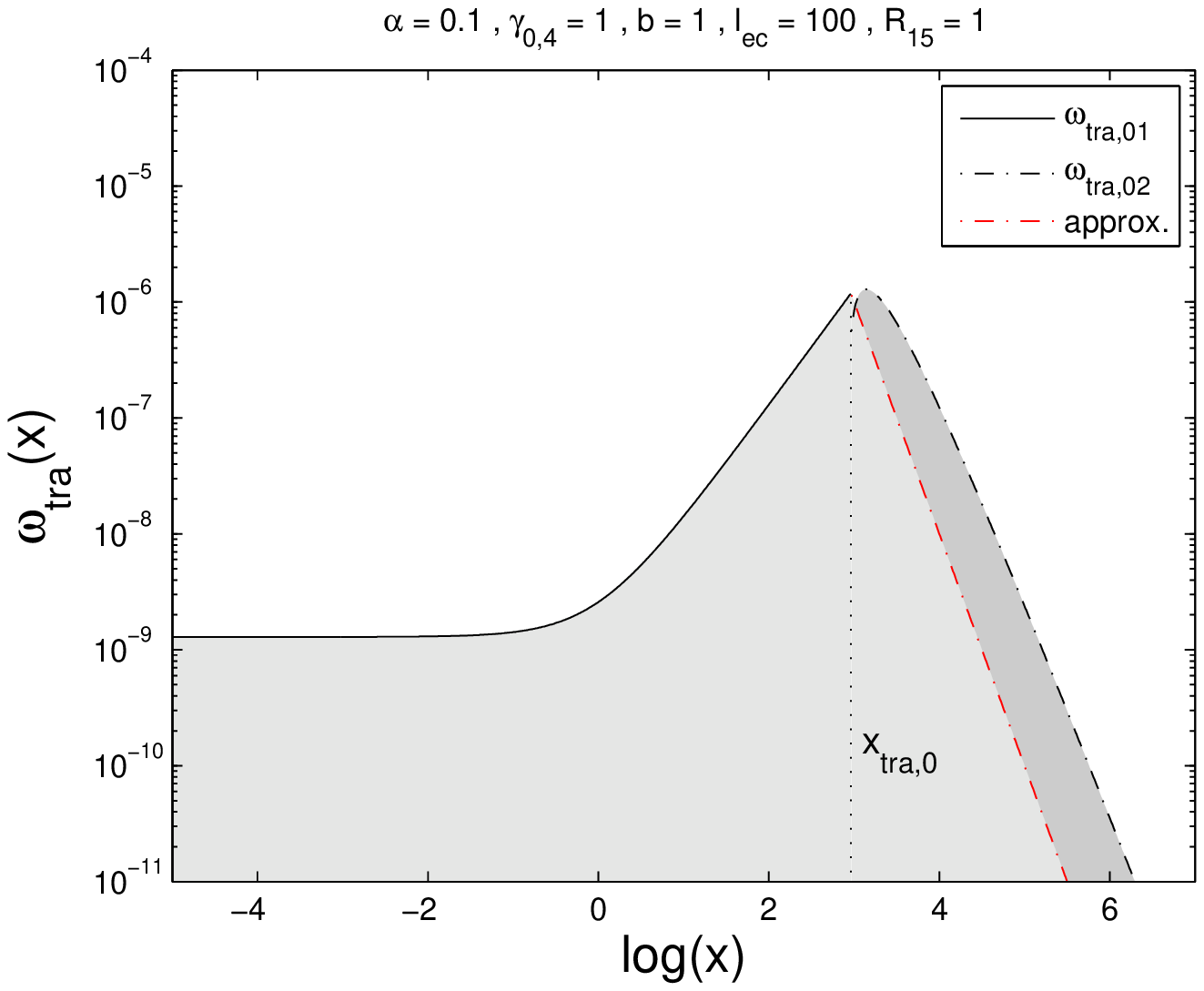}
	\caption{Transition energy $\omega_{tra}$ as a function of time $x$ for $\alpha<1$. The black solid curve shows $\omega_{tra,01}$, the black dot-dashed curve $\omega_{tra,02}$, and the red dot-dashed curve the approximation of $\omega_{tra,02}$. The transition time $x_{tra,0}$ is indicated by the black vertical dashed line. The light grey area marks the optical thick regime bounded by the approximation, while the dark area marks the optical thick regime bounded by the correct value of $\omega_{tra,02}$. Parameters are given at the top.}
	\label{fig:ssa01}
\end{figure}

In the opposite case $\eps(1+x)^2/\epsn\gg 1$, stronger approximations are necessary:
\begin{align}
1 \approx& \frac{3}{2}I_{ssa} \omega_{tra,02}^{-2/3} \left( 1+x \right)^{11/3} \nonumber \\
&\times e^{-\omega_{tra,02}(1+x)^2} \nonumber \\
\Leftrightarrow \frac{3}{2} I_{ssa} \left( 1+x \right)^5 =& \left[ \omega_{tra,02} \left( 1+x \right)^2 \right]^{2/3} e^{\omega_{tra,02}(1+x)^2} \eqd
\end{align}
The right-hand-side of this equation can be approximated as follows
\begin{align}
& \left[ \omega_{tra,02} \left( 1+x \right)^2 \right]^{2/3} e^{\omega_{tra,02}(1+x)^2} \nonumber \\
= & e^{\omega_{tra,02}(1+x)^2 + \frac{3}{2}\ln{\left[ \omega_{tra,02} \left( 1+x \right)^2 \right]}} \nonumber \\
\approx & e^{\omega_{tra,02}(1+x)^2} \eqc
\end{align}
giving
\begin{align}
\omega_{tra,02} &\approx \frac{1}{\left( 1+x \right)^2} \ln{\left[ \frac{3}{2}I_{ssa}\left( 1+x \right)^5 \right]} \nonumber \\
&\approx \left( 1+x \right)^{-2} \label{eq:epstra01b} \eqd
\end{align}
The quality of the last approximation can be checked in figure \ref{fig:ssa01}.

The transition time from $\omega_{tra,01}$ to $\omega_{tra,02}$ is given by
\begin{align}
x_{tra,0} &= \omega_{tra}^{-1/2}-1 \nonumber \\
\Rightarrow x_{tra,0} &= I_{ssa}^{-1/5}-1 \label{eq:xtra0} \eqd
\end{align}
For $I_{ssa}>1$ the transition time is negative, and only the second transition energy needs to be considered. However, as stated above, this case should not occur under normal circumstances.

As is obvious from figure \ref{fig:ssa01}, the transition energy depends strongly on time. Hence, the optical depth of the source with respect to synchrotron-self absorption depends on time, as well. The maximum energy is attained at the transition time with an energy $\epstra{ssa,max} = I_{ssa}^{2/5}\epsn = 2.5\cdot 10^{-11} \alpha^2 / \gamf^5$. In the example of figure \ref{fig:ssa01} this corresponds to a wavelength on the order of a few meter.
%
\subsection{The case $\alpha>1$} \label{app:ssa10}
For the initial SSC cooling case $\alpha>1$, equations (\ref{eq:eldis1}) and (\ref{eq:eldis2}) must be used in equation (\ref{eq:ssagen1}), giving
\begin{align}
\taus(\eps,x<x_c) = &I_{ssa} \left( \frac{\eps}{\epsn} \right)^{-5/3} \left( 1+3\alpha^2x \right)^{5/9} \nonumber \\
&\times \left[ 1+\frac{3}{2}\frac{\eps}{\epsn}\left( 1+3\alpha^2x \right)^{2/3} \right] e^{-\frac{\eps}{\epsn}\left( 1+3\alpha^2x \right)^{2/3}} \label{eq:taus10a} \eqc 
\end{align}
and
\begin{align}
\taus(\eps,x>x_c) =& I_{ssa} \left( \frac{\eps}{\epsn} \right)^{-5/3} \left( \alpha_g+x \right)^{5/3} \nonumber \\
&\times \left[ 1+\frac{3}{2}\frac{\eps}{\epsn}\left( \alpha_g+x \right)^2 \right] e^{-\frac{\eps}{\epsn}\left( \alpha_g+x \right)^2} \label{eq:taus20a} \eqc 
\end{align}
respectively.

Approximating as in the previous section yields obviously 4 transition energies. Using the steps outlined in the previous section, one obtains for $x<x_c$
\begin{align}
\omega_{tra,11} &= I_{ssa}^{3/5} \left( 1+3\alpha^2x \right)^{1/3} \label{eq:epstra10a} \eqc \\
\omega_{tra,12} &= \frac{1}{\left( 1+3\alpha^2x \right)^{2/3}} \ln{\left[ \frac{3}{2}I_{ssa} \left( 1+3\alpha^2x \right)^{5/3} \right]} \nonumber \\
&\approx \left( 1+3\alpha^2x \right)^{-2/3} \label{eq:epstra10b} \eqd
\end{align}
The transition time becomes
\begin{align}
x_{tra,1} &= \frac{1}{3\alpha^2}\left( \omega_{tra,1}^{-3/2} -1 \right) \nonumber \\
\Rightarrow x_{tra,1} &= \frac{1}{3\alpha^2}\left( I_{ssa}^{-3/5} -1 \right) \label{eq:xtra1} \eqd
\end{align}

For $x>x_c$ similar steps lead to
\begin{align}
\omega_{tra,21} &= I_{ssa}^{3/5} \left( \alpha_g+x \right) \label{eq:epstra10c} \eqc \\
\omega_{tra,22} &= \frac{1}{\left( \alpha_g+x \right)^2} \ln{\left[ \frac{3}{2}I_{ssa}\left( \alpha_g+x \right)^5 \right]} \nonumber \\
&\approx \left( \alpha_g+x \right)^{-2} \label{eq:epstra10d} \eqd
\end{align}
Here, the transition time is found to be
\begin{align}
x_{tra,2} &= \omega_{tra}^{-1/2}-\alpha_g \nonumber \\
\Rightarrow x_{tra,2} &= I_{ssa}^{-1/5}-\alpha_g \label{eq:xtra2} \eqd
\end{align}

Relating both transition times to $x_c$ like $x_{tra,1,2}>x_c$, both cases result in the same relation: $I_{ssa}^{-1/5}>\alpha$, implying $\alpha<67.7\gamf b^{1/7}$. Hence, under normal circumstances ($\alpha\sim 10$) one obtains $x_{tra,1}>x_c$, and therefore the transition energy changes directly from $\omega_{tra,11}$ to $\omega_{tra,21}$. In the opposite case, the transition would occur from $\omega_{tra,12}$ to $\omega_{tra,22}$.

\begin{figure}
	\centering
		\includegraphics[width=0.48\textwidth]{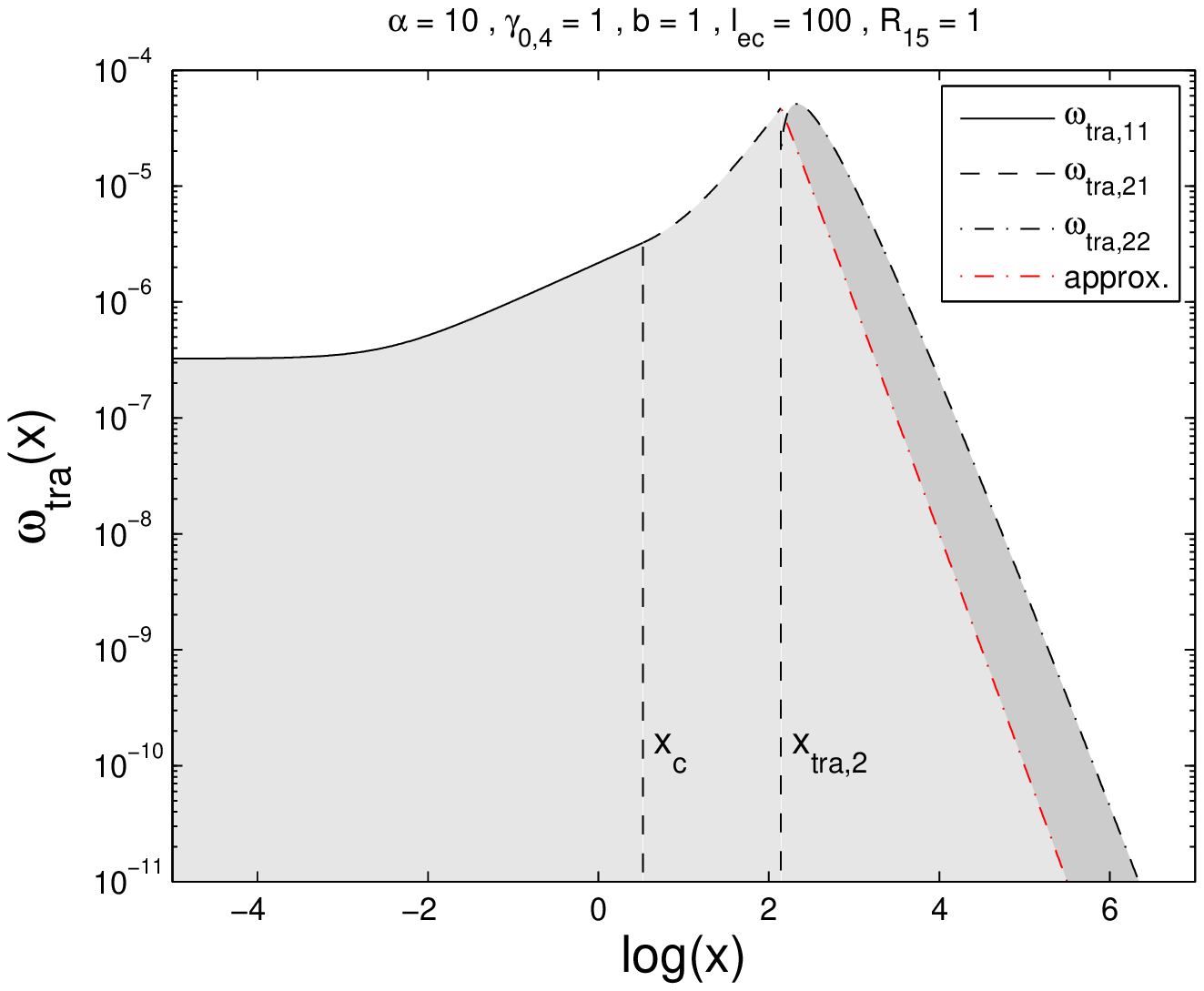}
	\caption{Transition energy $\omega_{tra}$ as a function of time $x$ for $\alpha>1$. The black solid curve shows $\omega_{tra,11}$, the black dashed curve $\omega_{tra,21}$, the black dot-dashed curve $\omega_{tra,22}$, and the red dot-dashed curve the approximation of $\omega_{tra,22}$. The transition times $x_c$, and $x_{tra,2}$ are indicated by the black vertical dashed lines. The light grey area marks the optical thick regime bounded by the approximation, while the dark area marks the optical thick regime bounded by the correct value of $\omega_{tra,02}$. Parameters are given at the top.}
	\label{fig:ssa10}
\end{figure}

The behaviour of the optical depth is similar to the $\alpha<1$ case, apart from the fact that more subcases are considered, as is obvious in figure \ref{fig:ssa10}. The maximum energy is the same as in the previous case $\epstra{ssa,max} = I_{ssa}^{2/5}\epsn = 2.5\cdot 10^{-11} \alpha^2 / \gamf^5$, which corresponds in the given example to a wavelength in the sub-mm regime.
%
\subsection{Remarks} \label{app:ssarem}
The time dependence of the optical depth implies some interesting features, which are not possible in a steady source. 

The general behaviour of the transition energies are rather similar in both cases of $\alpha$. At first the optical depth increases and after reaching the maximum energy (which is the same function in all cases, but depends on the chosen parameters) it begins to decrease, again. For $x\rightarrow\infty$ the source becomes completely optically thin. Hence, in the total SEDs of figure \ref{fig:sed} the optically thick part of the synchrotron SED is not visible, since the integration is from zero to infinity. 

There are also important differences between the two cases of $\alpha$. For $\alpha>1$ the increase starts a factor $3\alpha^2$ earlier than in the $\alpha<1$ case, and the first increase is not as steep. As expected, they are the same beyond $x_c$. 

The SSA transition energy could be used as a proxy for the source parameters (e.g., \citet{nbs14}), since in steady-state scenarios the transition energy is constant. However, in the given time-dependent injection scenario the parameters cannot be derived from the transition energy as easily, unless the observation time with respect to the onset of the flare (i.e., particle injection) is known. On the other hand, if the maximum transition energy can be deduced from the data, the parameter space can be significantly reduced, since $\epstra{ssa,max}$ depends only on 2 free parameters (3, if the Doppler factor is taken into account). 
%
%

%
%
%
\end{document}